\documentclass[journal=jctcce,manuscript=article,layout=twocolumn]{achemso}

\usepackage[version=3]{mhchem} 
\usepackage[T1]{fontenc}       

\usepackage{geometry}
\usepackage[utf8]{inputenc}
\usepackage{amsfonts,amsmath,amsbsy,color}
\usepackage{textcomp}
\usepackage[version=3]{mhchem}
\usepackage{braket}
\usepackage{graphicx}
\usepackage{subfig}
\usepackage{caption}
\usepackage[export]{adjustbox}
\usepackage{gensymb}

\author{Kristopher T. Jensen}
\author{Raz L. Benson}
\author{Salvatore Cardamone}
\author{Alex J.W. Thom}
\email{ajwt3@cam.ac.uk}
\affiliation[University of Cambridge]
{Department of Chemistry, University of Cambridge, Cambridge}

\title[SCF ET]
{Modelling Electron Transfers Using Quasidiabatic Hartree-Fock States}

\begin{document}








\begin{abstract}
Electron transfer processes are ubiquitous in chemistry and of great importance in many systems of biological and commercial interest. The ab-initio description of these processes remains a challenge in theoretical chemistry, partly due to the high scaling of many post-Hartree--Fock computational methods. This poses a problem for systems of interest that are not easily investigated experimentally. We show that readily available Hartree--Fock solutions can be used as a quasidiabatic basis to understand electron transfer reactions in a Marcus framework. 
Non-orthogonal configuration interaction calculations can be used to quantify interactions between the resulting electronic states, and to investigate the adiabatic electron transfer process.
When applied to a titanium-alizarin complex used as a model of a Grätzel-type solar cell, this approach yields a correct description of the electron transfer and provides information about the electronic states involved in the process.
\end{abstract}

\section{Introduction}

Electron transfer is fundamental to many chemical processes, and its description has long been a challenge in theoretical chemistry.
The work of Marcus\cite{Marcus1964}, and later refinements thereof\cite{Richardson2015semiclassical}, provides the framework for understanding such processes in terms of diabatic electronic states of acceptor and donor complexes, but these idealized viewpoints have proven very challenging to tackle from an ab initio electronic structure perspective, whose natural domain is to produce the adiabatic states which minimize the electronic energy.\\

A strict electronic basis of diabatic states can be shown to result when the nuclear derivative coupling is zero. Such a basis has been shown in general not to exist\cite{Mead1982}. Construction of the diabatic states (`diabatization') through minimisation of the vibronic coupling between basis states\cite{Baer1980} or analysis of configuration interaction expansions\cite{Atchity1997determination,Nakamura2003extension} is complicated by the significant computational effort required.\\

A wealth of more computationally tractable methodologies have been devised for approximating the diabatic states of a molecular system; block diagonalisation\cite{Pacher1993adiabatic, Domcke1994diabatic} and variants of the generalised Mulliken-Hush (GMH) algorithm\cite{Cave1997calculation,Subotnik2008constructing} have proven popular. More recently, constrained DFT (CDFT)\cite{Wu2006extracting, Voorhis2010} and its coupling with configuration interaction (CDFT-CI)\cite{Wu2007configuration} for treating strongly correlated systems have been proposed. However, both CDFT and the GMH algorithm require some imposed intuition to define localised charges, rendering the resultant diabatic states dependent upon the charge localisation scheme invoked. CDFT-CI has also been shown occasionally to fail significantly\cite{Kubas2014electronic}, although a metric has recently been developed to predict a system's propensity for poor delineation\cite{Mavros2015communication}. Furthermore, the excited states of certain classes of molecules, such as cyanide dyes and retinal chromophores, are poorly characterised by virtually all DFT functionals\cite{Demoulin2016intramolecular}, restricting the domain of application for such techniques.\\

Recently, one of us has shown \cite{Thom2009} that a simple alternative to diabatization is to use the multiple solutions of the Hartree--Fock equations, which behave as quasi-diabatic states, and has given a simple methodology for locating them\cite{Thom2008}.
These states have been found to yield qualitatively accurate excited state orbitals and approximate single and double vertical excitation energies \cite{Sundstrom2014}. They exhibit a lack of avoided crossings and provide an approximately constant electronic structure across a wide range of geometries,  as expected for diabatic states.\cite{Thom2009} \\

These excited SCF solutions were also used as a basis for non-orthogonal configuration interaction (NOCI) by solving the generalized eigenvalue problem
\begin{equation}\label{eq:NOCI}
\mathbf{H D} = \mathbf{S D E}
\end{equation}
where $\mathbf{H}$ is the Hamiltonian matrix, $\mathbf{E}$ the diagonal energy matrix, and $\mathbf{S}$ the overlap matrix. The vectors which form $\mathbf{D}$ describe the resulting NOCI states in terms of linear combintations of the SCF solutions.
These NOCI states exhibit avoided crossings and conical intersections as expected for adiabatic electronic states, and also resolve heavily spin-contaminated SCF solutions into their constituent spin-pure states\cite{Thom2009}.\\

While investigating the multiple Hartree--Fock solutions, Thom and Head-Gordon found that some of the excited SCF states coalesce and disappear from conventional Hartree--Fock space. This is similar to the well-known Coulson-Fischer point\cite{CoulsonFischer} of \ce{H2}, where the two ground state UHF solutions coalesce as the UHF solutions become identical to each other and the RHF solution. 
To investigate this disappearence of SCF solutions from conventional Hartree--Fock space, Hiscock and Thom developed the method of Holomorphic Hartree--Fock Theory, in which complex holomorphic UHF (hUHF) solutions are found as stationary points of a holomorphic energy functional \cite{Hiscock2014}.
Burton and Thom later showed that holomorphic SCF states can be used together with real SCF states to form a continuous basis for NOCI, allowing coherent descriptions of processes such as molecular deformations and bond-breaking across geometries where SCF solutions coalesce and vanish\cite{Burton2016}.\\

In the present paper, we further investigate the topics of multiple and coaslescing SCF states by characterizing excited SCF solutions for two medium-sized systems and investigating the physical implications of their coaslescence.
We proceed to show how such excited SCF solutions can be used to model electron transfer processes in a diabatic framework, avoiding the need for conventional diabatization methods. Finally, we apply this methodology to an alizarin-titanium complex of interest in the rapidly growing field of dye-sensitized solar cells and recover a model of the electron transfer process consistent with previous experiments.

\section{Results and Discussion}

\subsection{A Model of Electron Transfer}
\label{ss:C7H6F4}

\begin{figure}[h]
\centering
 \subfloat{\includegraphics[width=0.7 \linewidth]{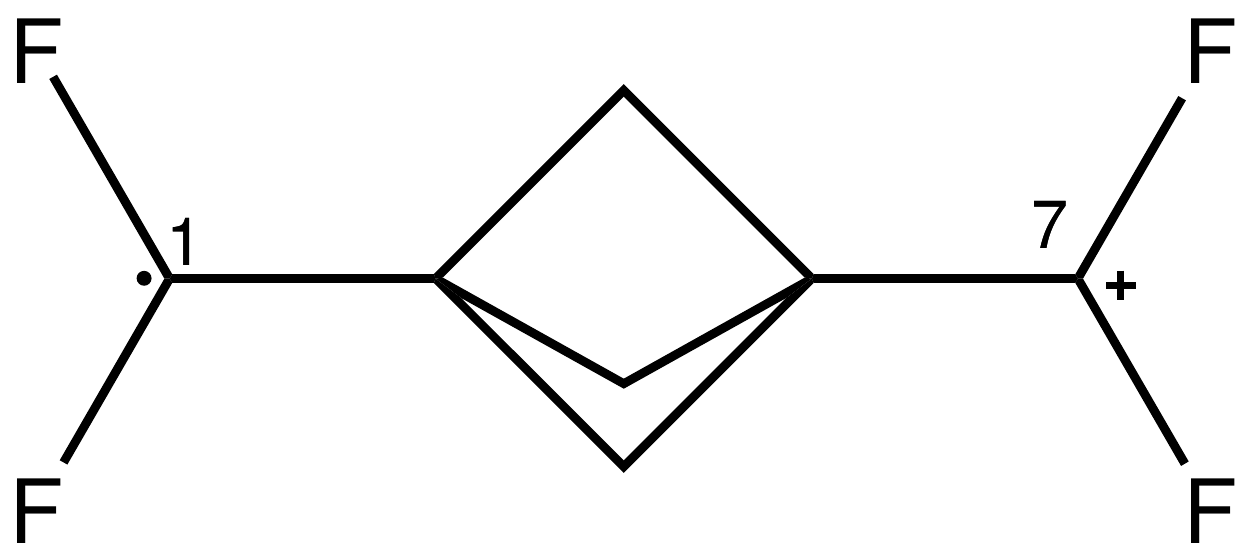}}
 \caption{Structure of 1,3-di-difluoromethylene-bicyclo[1.1.1.]pentane (\ce{C7H6F4^.+}). C1 (left) is the electron donor and C7 (right) the electron acceptor}
 \label{fig:structure}
\end{figure}

In order to model the behavior of SCF solutions in a relatively simple electron transfer process, we investigate the radical cationic doublet state of the bicyclo[1.1.1.]pentane-derivative \ce{C7H6F4^.+} shown in Figure \ref{fig:structure}. This is a tetra-fluoride derivative of a model system previously used for studies on the effect of physical donor-acceptor separation on electron transfer rates and coupling elements. \cite{Liang1992, Pati2002, Mo2003}. Inclusion of the fluorides increases the stability of the localized electron and simplifies identification and tracking of the relevant electronic states.\\


For this system, we investigate the transition of a single electron from one terminal \ce{CF2} group (C1) to the opposite terminal \ce{CF2} group (C7) working in Dunning's cc-pVDZ basis \cite{Dunning1989}.
Two degenerate SCF ground states exist for a symmetrical geometry optimized for the neutral singlet molecule. Inspection of the Mulliken charges confirm the two states to be symmetry-broken mirror images of one another, and the singly occupied Boys' localized orbitals identify the two states as corresponding to the donor (D) and acceptor (A) electronic states involved in the electron transfer with a radical electron on C1 and C7 respectively.\\

Each of these two states is used for an SCF geometry optimization which leads to the equilibrium geometries for the proposed  D and A states, both with ground-state electronic energies of $ - 665.92098 \, \mathrm{E_h} $ due to the symmetry of the system. In the following, all energies for \ce{C7H6F4^.+} are given relative to this energy.
In order to approximate the transition state of the electron transfer, we identify the Minimum Energy Crossing Point (MECP) of the two SCF solutions\cite{Koga1985} with energy  $0.01973 \, \mathrm{E_h} $. The MECP is defined as the lowest energy geometry for which the D and A states are degenerate, and we expect this to represent an energy maximum of the reaction trajectory.
An SCF metadynamics calculation performed for the MECP geometry identifies an additional low-lying symmetric state (E) with an energy of $ 0.03264 \, \mathrm{E_h} $ which we hypothesize can be used together with the donor and acceptor states as a basis for NOCI along the reaction trajectory.
\\

\begin{figure}[h]
\centering
 \subfloat{\includegraphics[width=\linewidth]{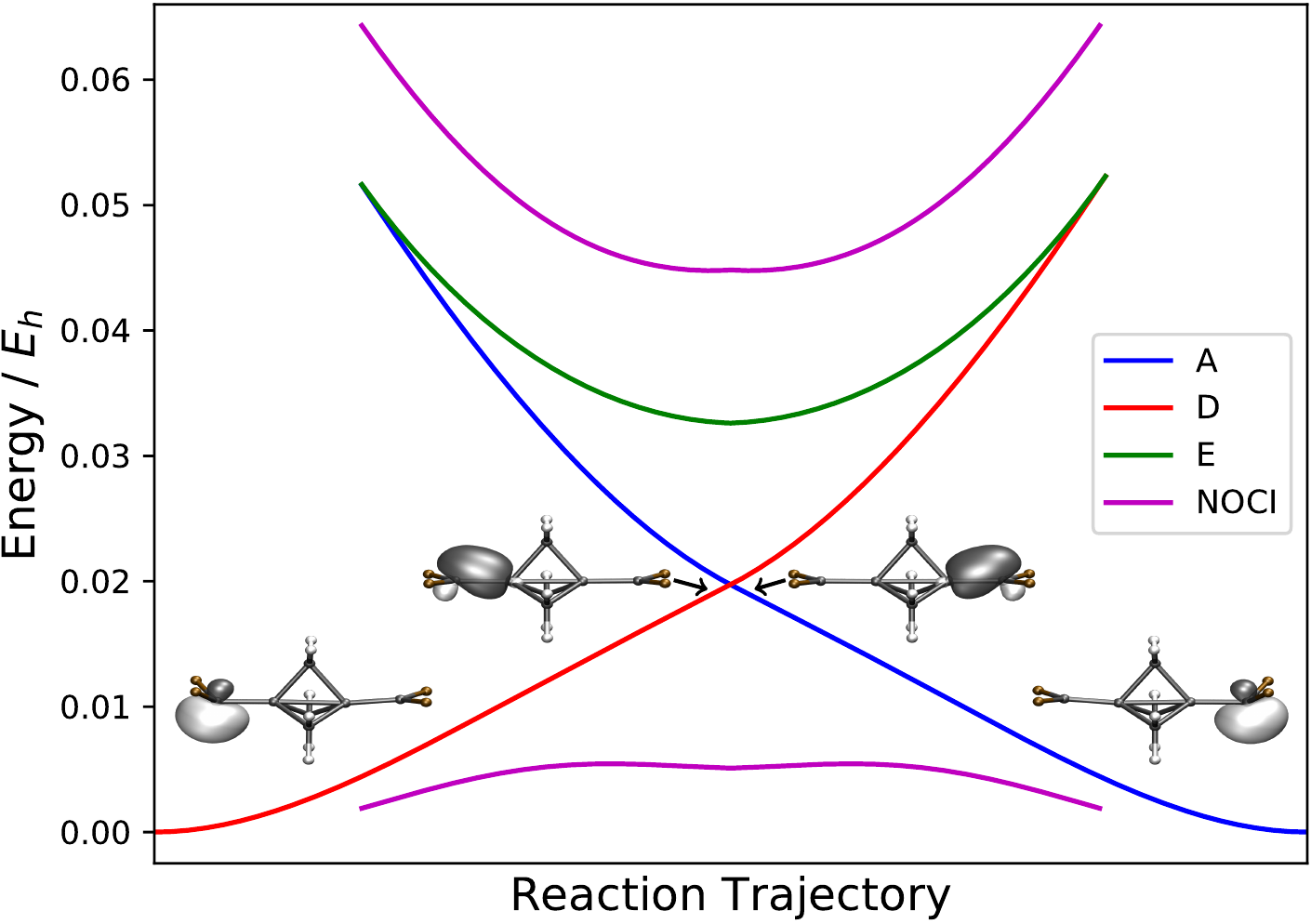}}

\caption{Energies of the D, A and E SCF states and two lowest energy NOCI states. The third NOCI state has an energy above 0.45 $\mathrm{E_h}$ and is not shown. \textit{Insets}: Singly occupied Boys' localized orbitals for the donor electronic state at the donor equilibrium geometry and MECP, and the acceptor electronic state at the acceptor equilibrium geometry and MECP. }

\label{fig:eclipsed_E}

\end{figure}

A linear interpolation between Z-matrices corresponding to the transition state and donor/acceptor geometries (available in supporting information) is used as an approximation to a reaction trajectory.
SCF metadynamics calculations are performed at each of 200 geometries along the trajectory and the electronic states of interest identified. The energies of the A, D and E states along this trajectory are given in Figure \ref{fig:eclipsed_E} together with the singly occupied Boys' localized orbitals of the D and A states.\\

In Figure \ref{fig:eclipsed_E}, it is interesting to note that the D electronic state coalesces with the E state 65 \% of the way from the MECP to the acceptor equilibrium geometry and vice versa for the A state. This is very similar to what has previously been observed for simpler molecules \cite{Hiscock2014, Burton2016, Thom2009}. It has been shown that the total number of holomorphic and non-holomorphic solutions must remain constant for two-electron systems, proving that the SCF solutions can be traced by holomorphic Hartree-Fock theory after coaslescence points\cite{Burton2018}. While this has not been rigorously proved for many-electron systems, we believe that it is theoretically be possible to follow the holomorphic SCF states throughout the reaction trajectory\cite{Burton2016}. This would provide a continuous basis for non-orthogonal configuration interaction to provide an approximation to a complete adiabatic energy profile.\\

While the coalescence of SCF states prevents us from calculating NOCI states that are continuous along the entire reaction trajectory, the three states are used as a basis for NOCI in the geometry space around the MECP, where all three SCF solutions exist. 
Near the MECP geometry, the NOCI ground state is significantly lower in energy than the D and A states, indicating that the donor and acceptor states are strongly interacting in this region with a lowering in energy (adiabatic correction factor\cite{Spencer2016}) of $0.0146 \, \mathrm{E_h}$. Strong donor-acceptor interactions are known to be associated with adiabatic electron transfers, \cite{Oberhofer2017} and indeed the NOCI ground state can be interpreted as an adiabatic description of the electron transfer.\\

\begin{figure}[h]
\centering
 \subfloat[][Ground NOCI state]{\adjincludegraphics[width=0.7 \linewidth,trim={0 {0.28 \height} 0  {0.29 \height}}, clip=true]{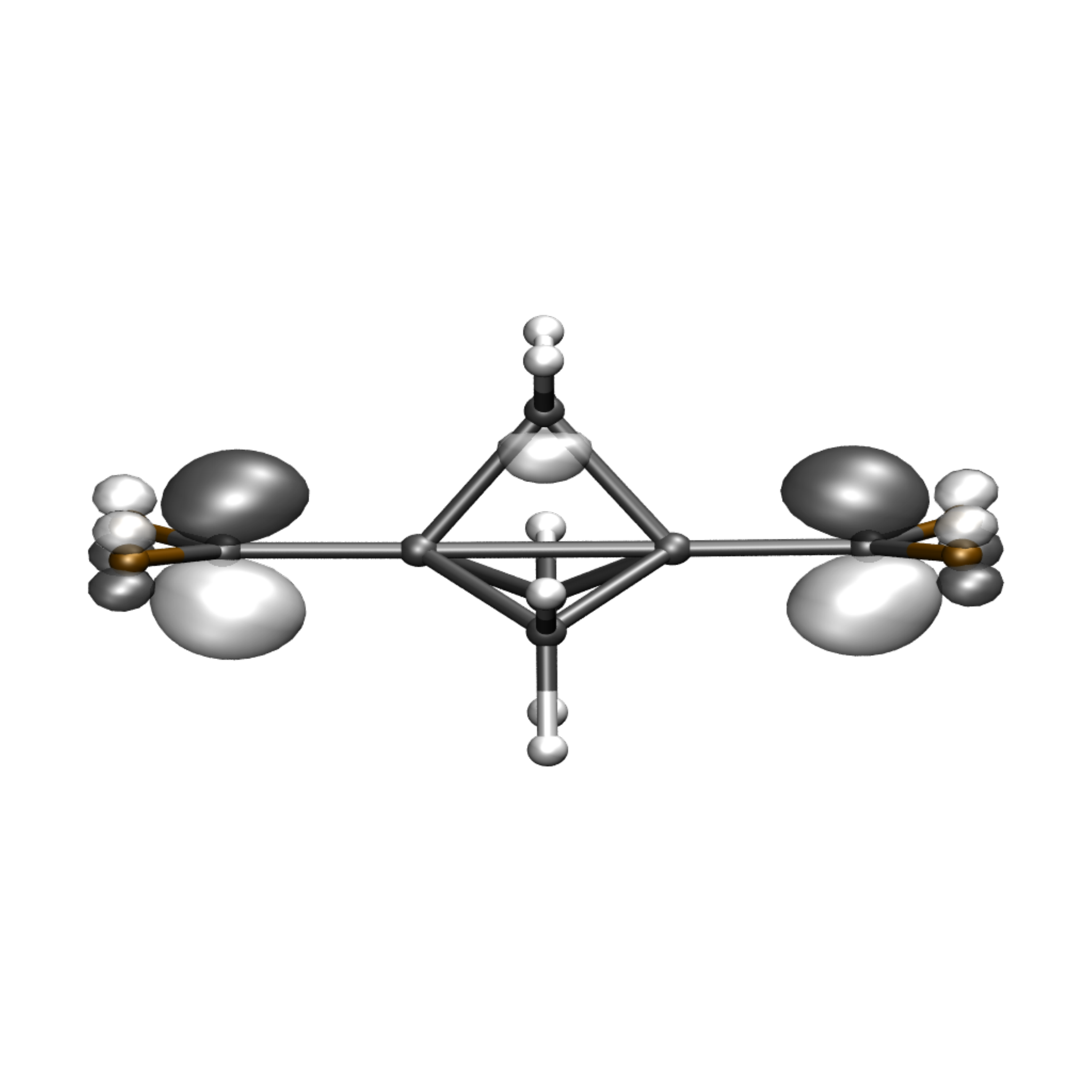}}\\
 \subfloat[][E SCF state]{\adjincludegraphics[width=0.7 \linewidth,trim={0 {0.28 \height} 0  {0.27 \height}}, clip=true]{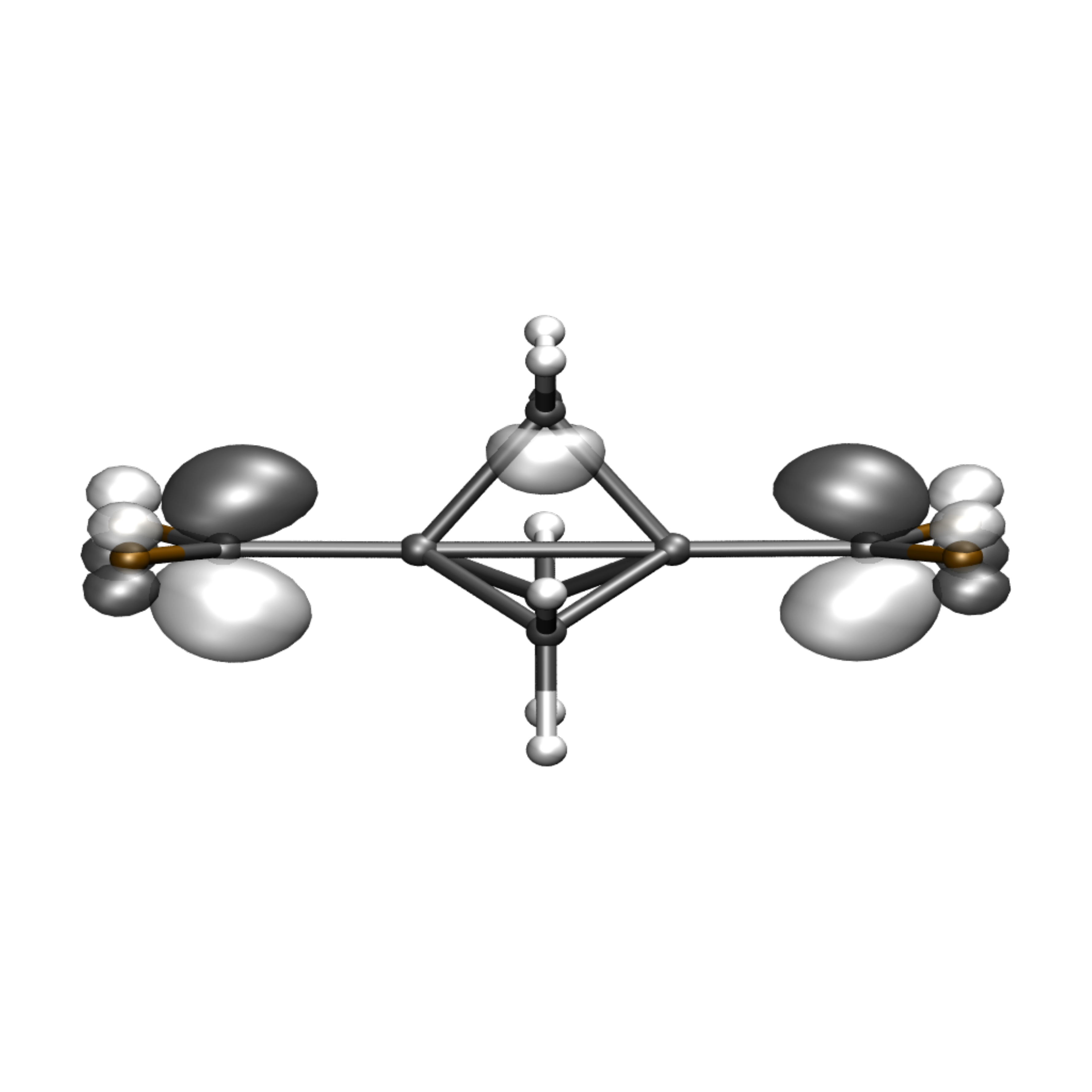}}\\
 \subfloat[][Excited NOCI state]{\adjincludegraphics[width=0.7 \linewidth,trim={0 {0.28 \height} 0  {0.27 \height}}, clip=true]{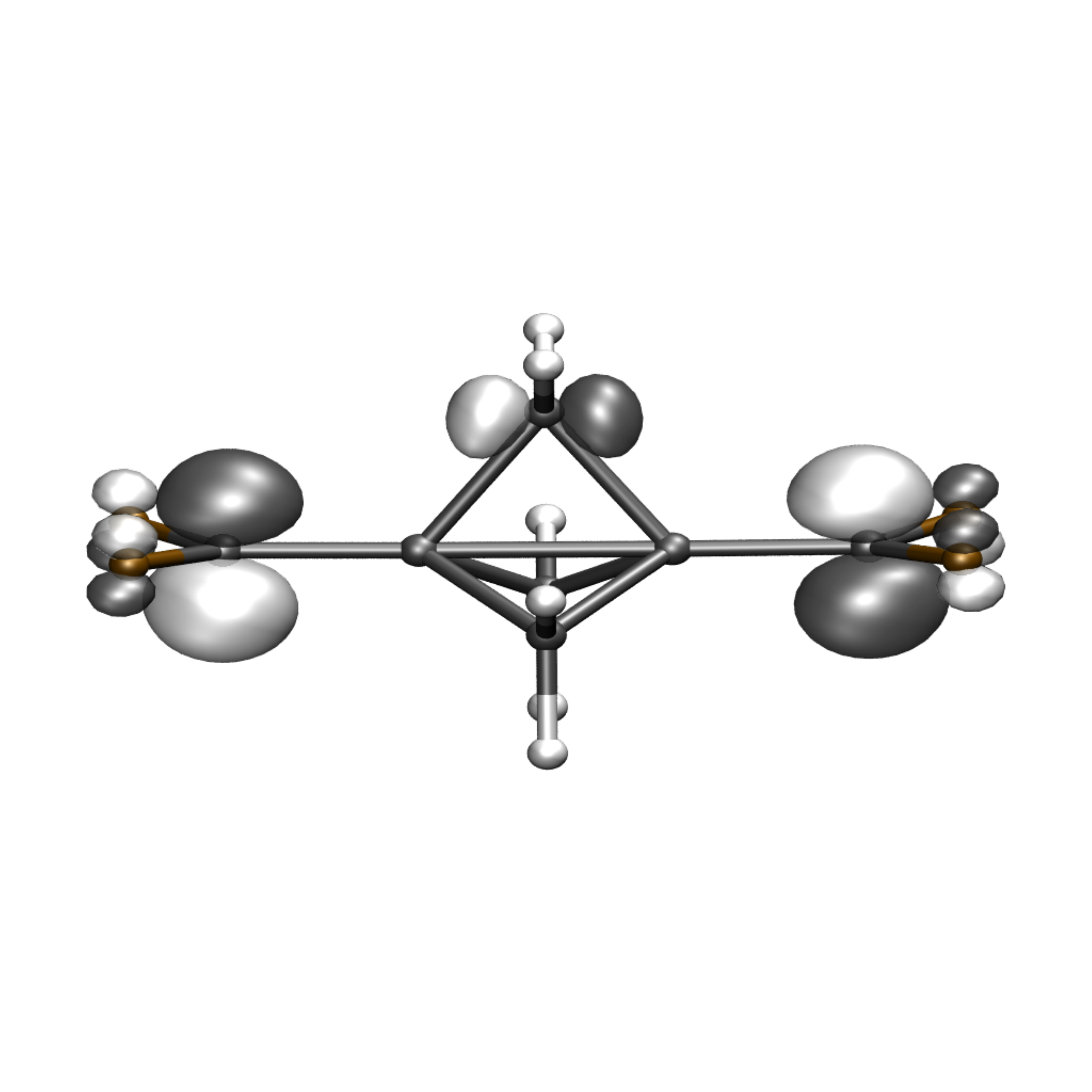}}
 \caption{Carbon-based natural orbitals (NOCI) and non-localized singly-occupied MO (SCF) at the MECP. The A and D SCF state singly occupied MOs are as shown in figure  \ref{fig:eclipsed_E}. }
 \label{fig:E_NOCI_orb}
\end{figure}

If the E state is not included in the NOCI calculation, the energy of the ground NOCI state is raised by less than $0.0014 \, \mathrm{E_h}$, indicating that inclusion of the E state does not extend the Hilbert space significantly.
In an attempt to understand the chemical significance of the E state, we can visualize the natural orbitals at the MECP of both the E SCF state and the NOCI ground and excited states, which correspond mostly to the in- and out-of-phase combinations respectively of the D and A states (Figure \ref{fig:E_NOCI_orb}).\\

These natural orbital plots indicate that despite being higher in energy than the D and A states, the E state is in fact very similar in terms of electron density to the NOCI ground state at the MECP geometry. This is supported by the electronic distance metric of Thom and Head-Gordon\cite{Thom2008}, extended to use the NOCI density matrices.  By this metric, the distance between E and the in-phase NOCI state is 0.056 electrons compared to a distance of 1.006 electrons between E and the higher energy out-of-phase NOCI state. At this MECP geometry, the E state can thus be interpreted as a single-determinant approximation to the delocalized NOCI ground state; the E state has a higher energy as it lacks electron correlation which is to some extent included in the NOCI ground state.\\

However, when moving further away from the MECP towards the donor or acceptor geometries, the electronic distance between E and the NOCI ground state increases to 0.744 at the point of E/D or E/A coalescence while the distance between E and the NOCI excited state decreases to 0.365. The E state is therefore a single-determinant approximation to a linear combination of the D and A states, but with the degree of in-phase compared to out-of-phase character changing along the reaction trajectory.\\

It is also informative to compare the Hartree-Fock derived quasidiabatic states to those obtained from a conventional CDFT calculation (using the B3LYP functional\cite{Becke93}) to see in what ways the two methods differ. In the CDFT calculations, the positive charge has been constained to be on C1 and the adjacent carbon and two fluorine atoms for the D state, and C7 and the adjacent atoms for the A state.\\

The Mulliken charges on the terminal carbon atoms for the Hartree-Fock states are 0.75 on C1 and 0.47 on C7 for the D state and vice versa for the A state. This can be compared to 0.47 and 0.19 for the CDFT states which thus show less charge localization. To offset the increased positive charge on the terminal carbon atoms, there is more negative charge associated with the central cage in the Hartree-Fock states compared to the CDFT states. In general, constrained DFT thus seems to spread out the charge more than the Hartree--Fock states for this system.\\

\begin{figure}[h]
\centering
 \subfloat[]{\adjincludegraphics[width=0.5 \linewidth,trim={0 {0.28 \height} 0  {0.29 \height}}, clip=true]{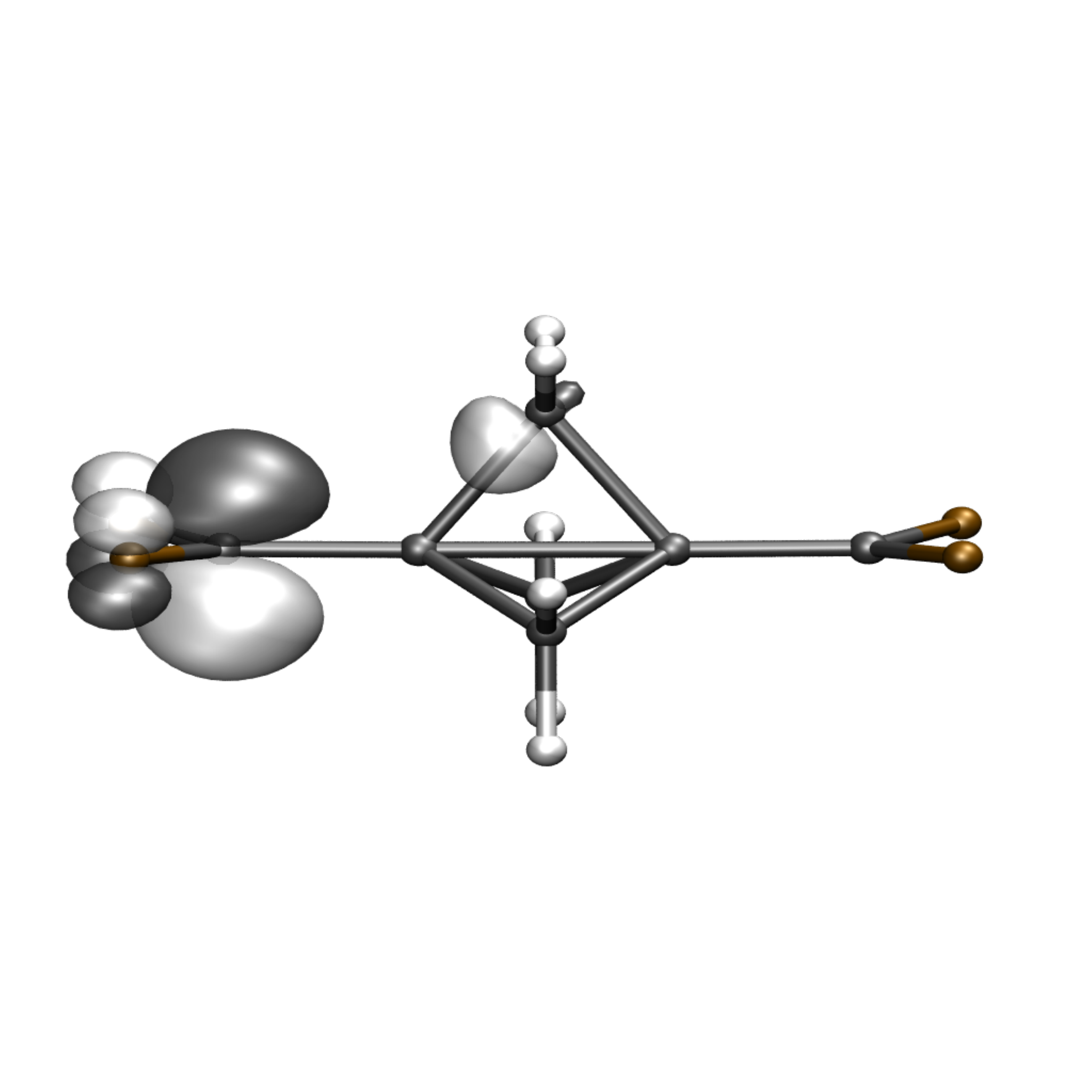}}
 \subfloat[]{\adjincludegraphics[width=0.5 \linewidth,trim={0 {0.28 \height} 0  {0.29 \height}}, clip=true]{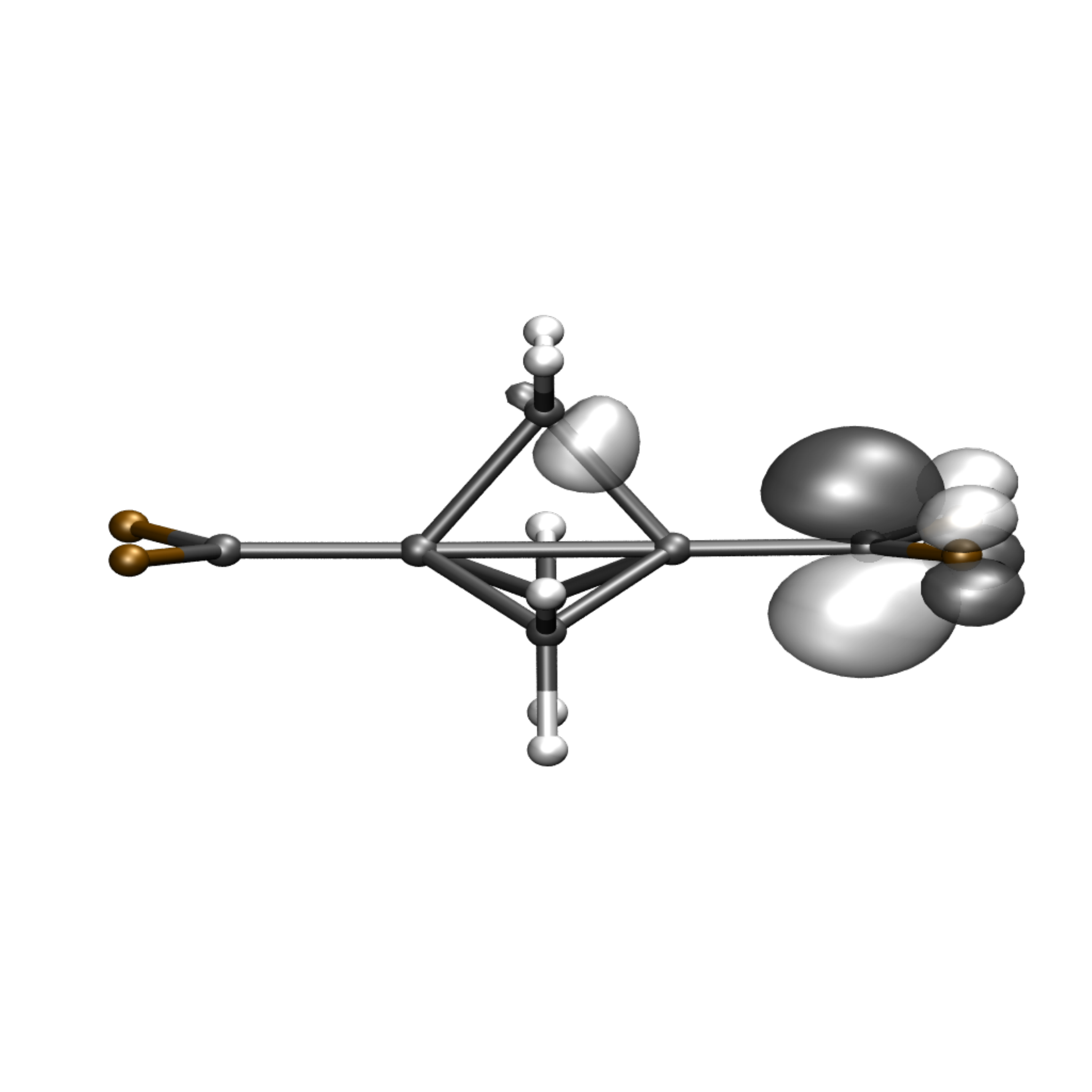}}\\
 \subfloat[]{\adjincludegraphics[width=0.5 \linewidth,trim={0 {0.28 \height} 0  {0.29 \height}}, clip=true]{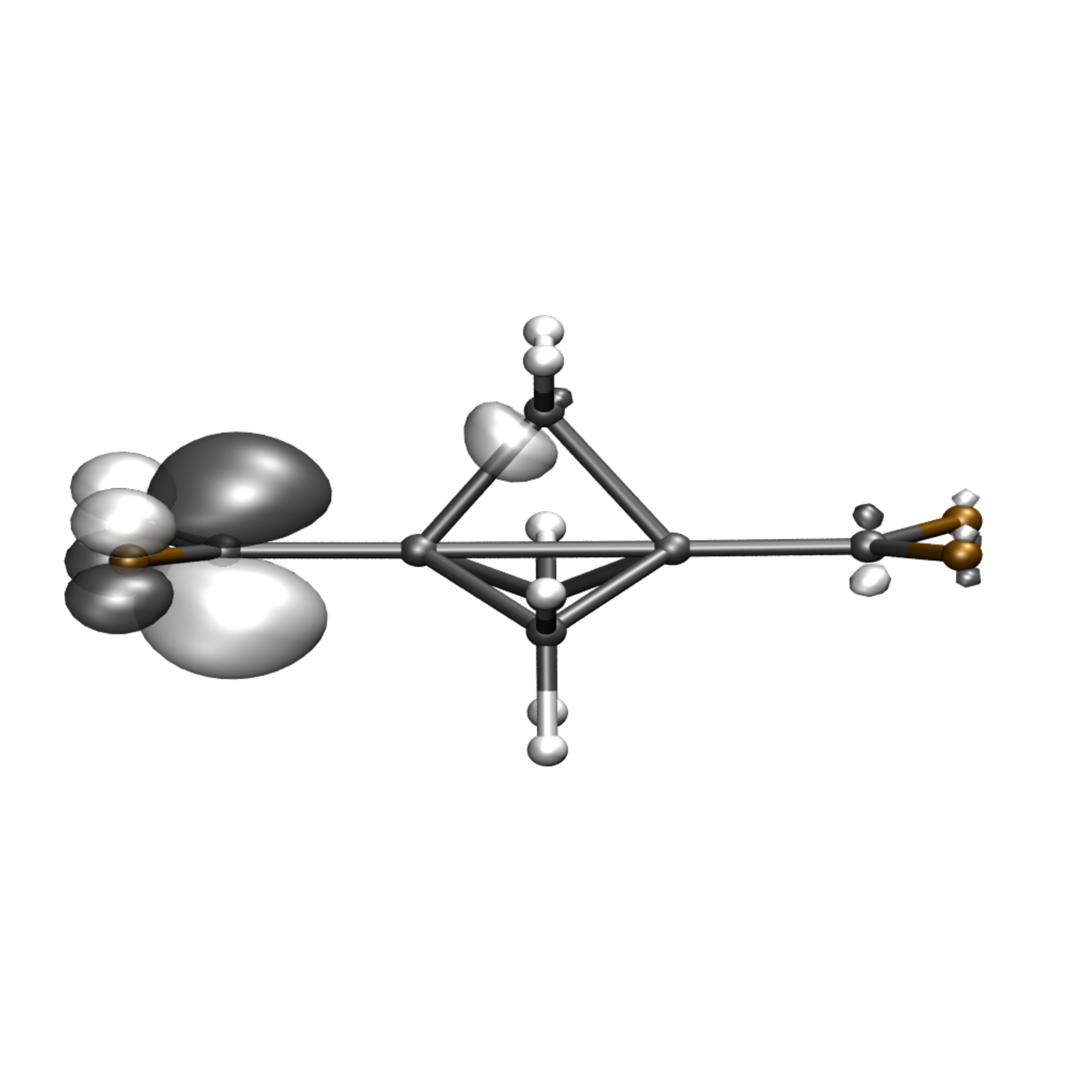}}
 \subfloat[]{\adjincludegraphics[width=0.5 \linewidth,trim={0 {0.28 \height} 0  {0.29 \height}}, clip=true]{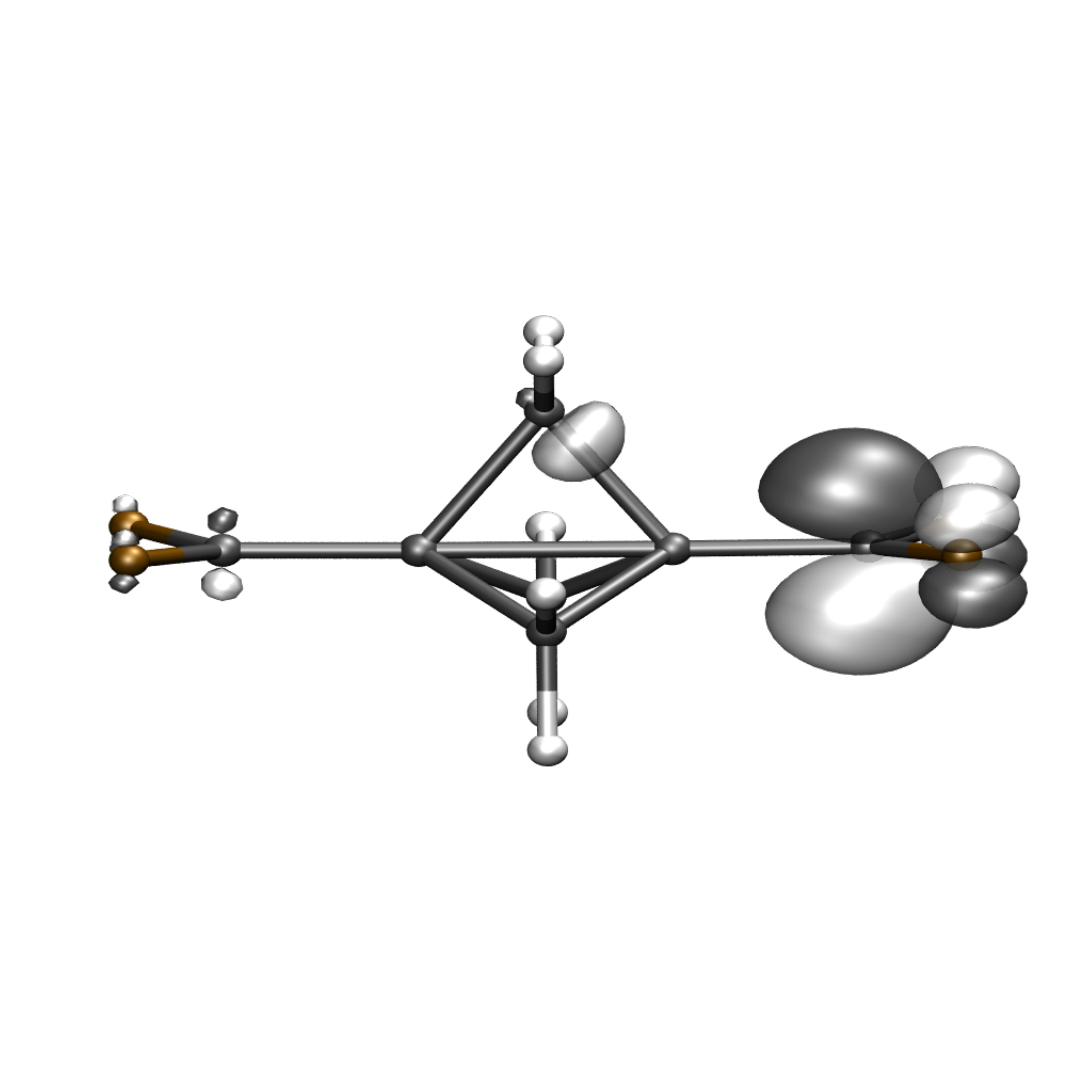}}
 \caption{Comparison of Hartree-Fock D (a) and A (b) states to the corresponding states obtained by CDFT (c and d) for the MECP geometry }
 \label{fig:cDFT}
\end{figure}

When progressing from the MECP to the donor equilibrium geometry, the HF Mulliken charges change to 0.73 and 0.56 for C1 and C7. CDFT gives Mulliken charges of  0.45 and 0.26 for this geometry. The change from the MECP to the donor geometry is thus similar between the two methods. It is also informative to compare the HOMOs from the two methods for the MECP geometry as illustrated in figure \ref{fig:cDFT}.

We see that CDFT and the quasidiabatic Hartree-Fock states give qualitatively very similar HOMOs. CDFT has the advantage that it is possible to ensure that a state with a given set of properties can be found by imposing appropriate constraints. However, CDFT also requires imposing subjective constraints on the system whereas the quasidiabatic Hartree--Fock states emerge directly from the SCF calculation. This provides a more unbiased approach to studying charge-localized states and electron transfers. It also facilitates the study of excited states which may be important in less simple systems with complex electronic structures and near-degenerate states. However, this comes at the cost that we cannot ensure the localization and identification of states, and that their coaslescence can hinder a complete picture of the electron transfer process.

\subsection{Two-Dimensional Energy Landscapes}
\label{ss:rs}

\begin{figure}[h]
\centering
 \subfloat{\includegraphics[width=0.7 \linewidth]{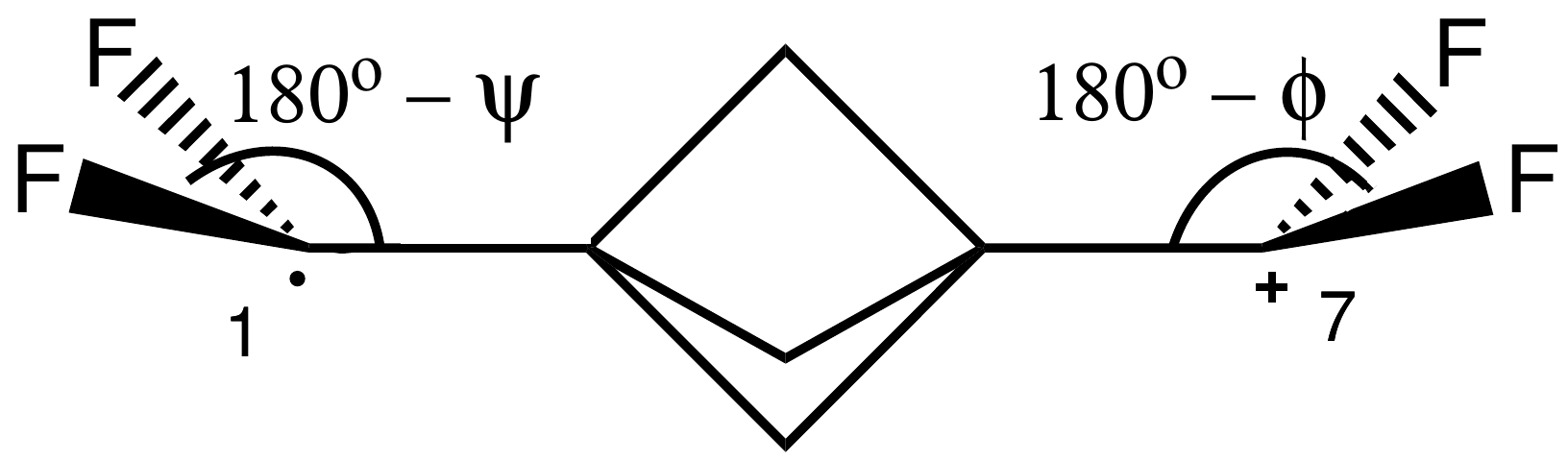}}
 \caption{Illustration of the parametization of the reaction space by the angle of the \ce{CF2} groups for \ce{C7H6F4^.+}}
 \label{fig:structure_ang}
\end{figure}

In order to characterize the behavior of the SCF solutions for \ce{C7H6F4^.+} more systematically and further illuminate the proposed electron transfer process, we systematically vary the $sp^3$-nature of the two terminal carbon atoms of the molecule. This is achieved by constructing 52 sets of CCF angles, FCF angles and CF bond lengths by interpolating between
\begin{enumerate}
 \item an optimized structure constrained to have a planar terminal \ce{CF2} group ($sp^2$), and
 \item the electron-containing terminal group of the donor geometry previously discussed ($sp^3$).
\end{enumerate}
A set of molecular geometries with combinations of these 52 terminal group conformations is generated and the energies and Mulliken charges of the D, A and E states calculated for each geometry. This parametization of the reaction space provides a simple visualization of the reaction on a two-dimensional energy surface as a function of the angles ($\phi, \psi$)  between each of the two \ce{CF2} planes and the central C-C-C-C axis (Figure \ref{fig:structure_ang}).\\

\begin{figure}[h]
\centering
 \subfloat{\adjincludegraphics[width=0.63 \linewidth, trim={  0 0 {0.05 \width}  0}, clip=true]{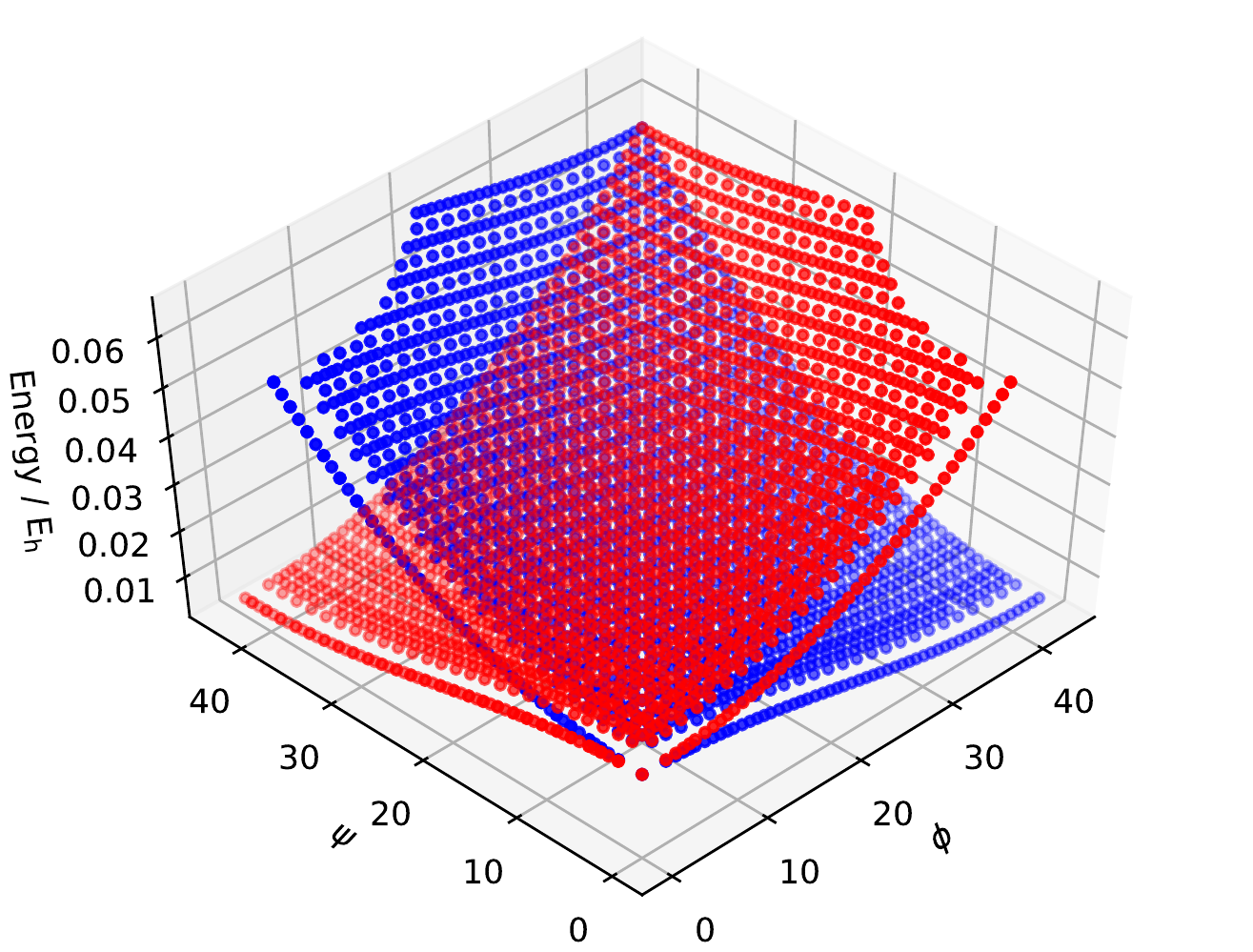}}
 \subfloat{\adjincludegraphics[width=0.59 \linewidth, trim={ {0.12 \width} 0 0  0}, clip=true]{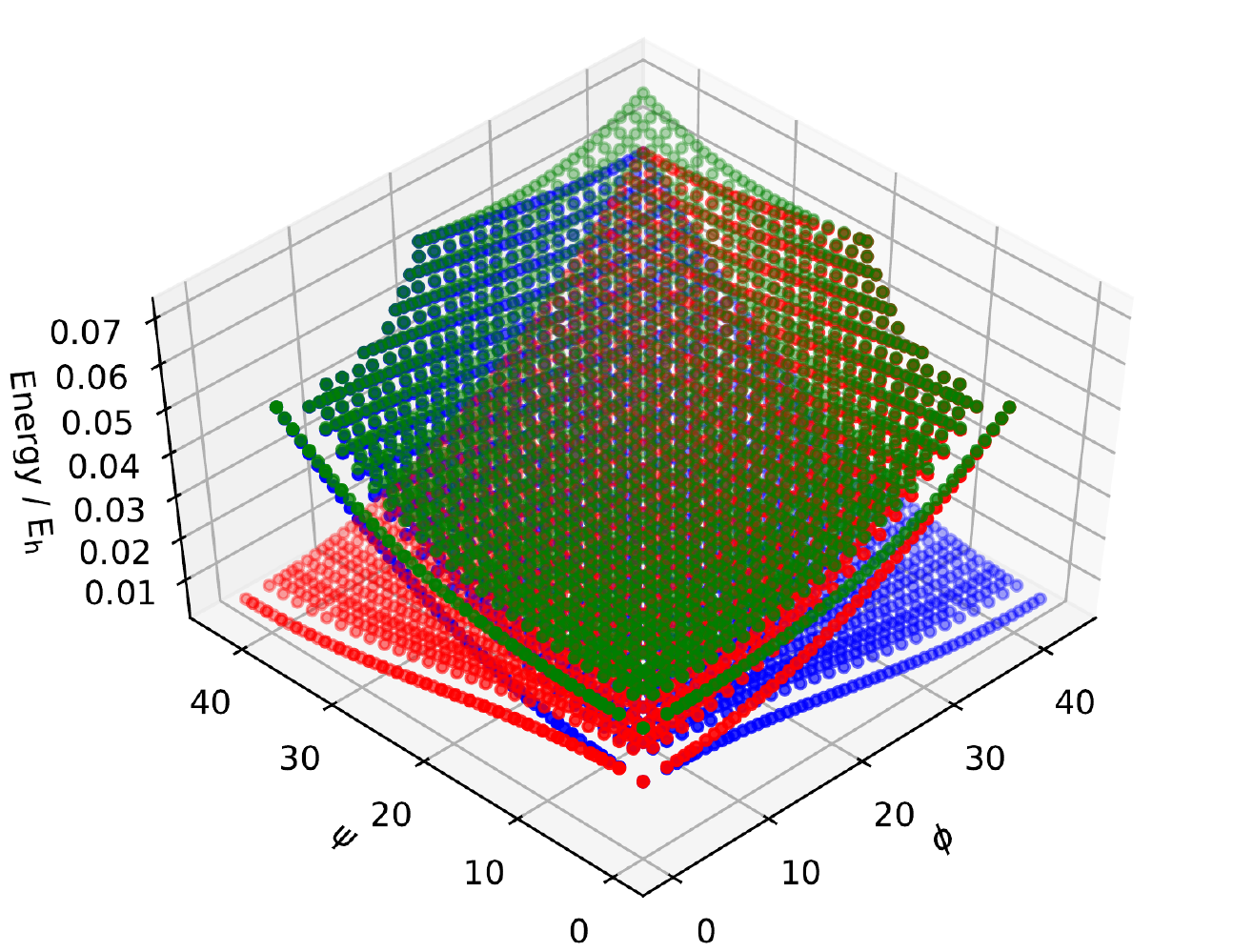}}\\
 \subfloat{\adjincludegraphics[width=0.8 \linewidth, trim={ 0 0 0 {0.10 \height}}, clip=true]{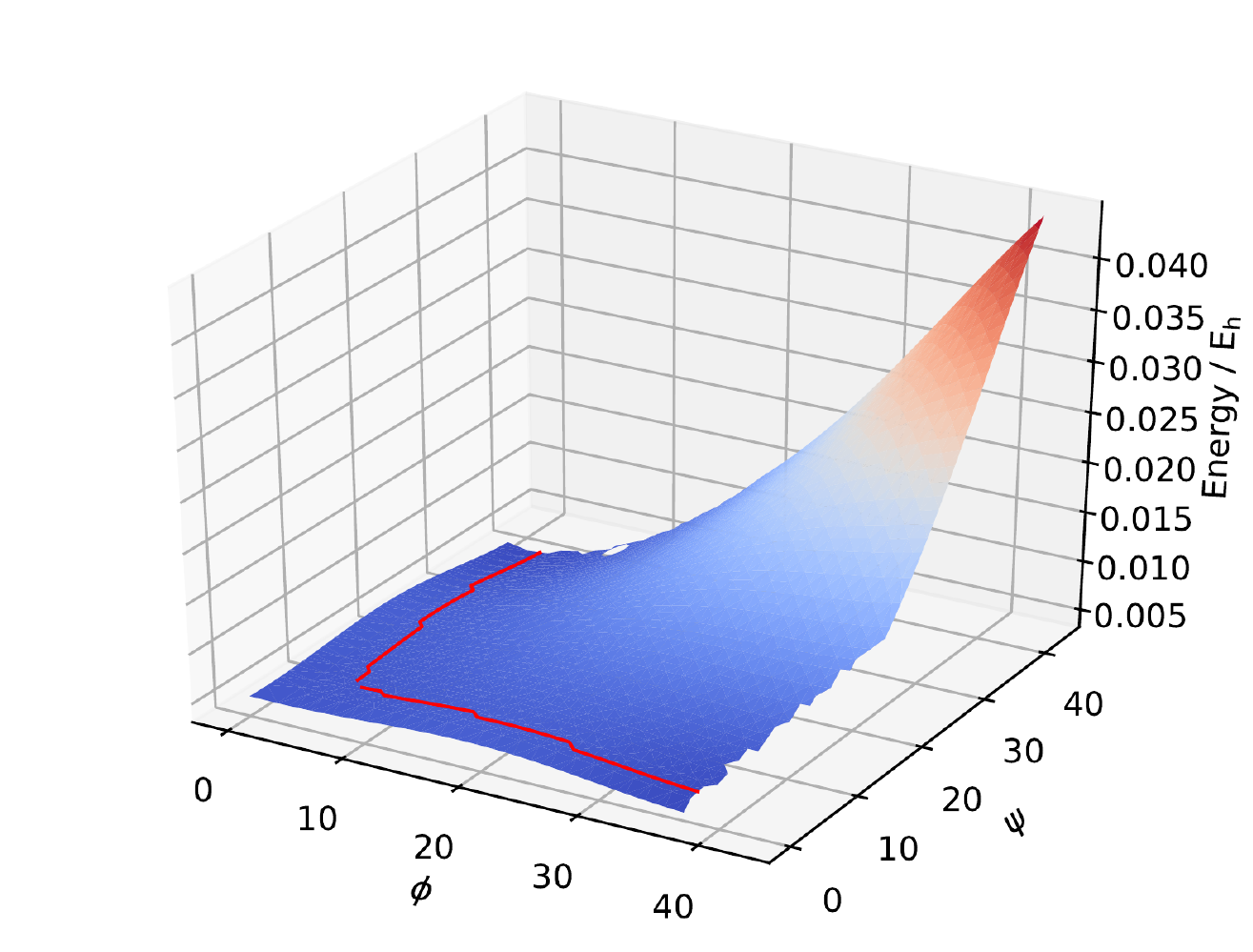}}
 \caption{\textit{top left:} Energy surfaces of the D and A SCF states as a function of $\phi$ and $\psi$. \textit{top right:} Energy surface of the E SCF state superimposed on the D and A surfaces illustrating coalescence of states. Red: D, Blue: A, Green: E. \textit{bottom:} Energy of the NOCI ground state (rotated relative to \textit{top}) as a function of $\phi$ and $\psi$. Red line: Reaction trajectory described in the previous section}
 \label{fig:C7H6F4_1_surfe}
\end{figure}

These energy surfaces address a number of interesting questions relating to the electron transfer process and the behavior of excited SCF solutions. Firstly, Figure \ref{fig:C7H6F4_1_surfe} gives a two-dimensional representation of the coalescence of the D and A states with the E state. It is evident that coalesence occurs when there is a significant difference between the $\phi$ and $\psi$ angles.
This is consistent with the interpretation of the E-state as a single-determinant approximation to a linear combination of D and A states since large $\phi$-$\psi$ separations represent the regime where the contribution of the lower energy SCF minimum to the out-of-phase NOCI state tends towards zero and the major contribution is from the higher energy SCF minimum.\\

Secondly, the energy surface for the NOCI ground state serves as a verification of our interpolation procedure as an approximation to the minimum energy reaction trajectory as can be seen from the proposed reaction trajectory remaining in a valley in the energy landscape.\\

\begin{figure}[h]
\centering
 \subfloat{\includegraphics[width=0.8 \linewidth]{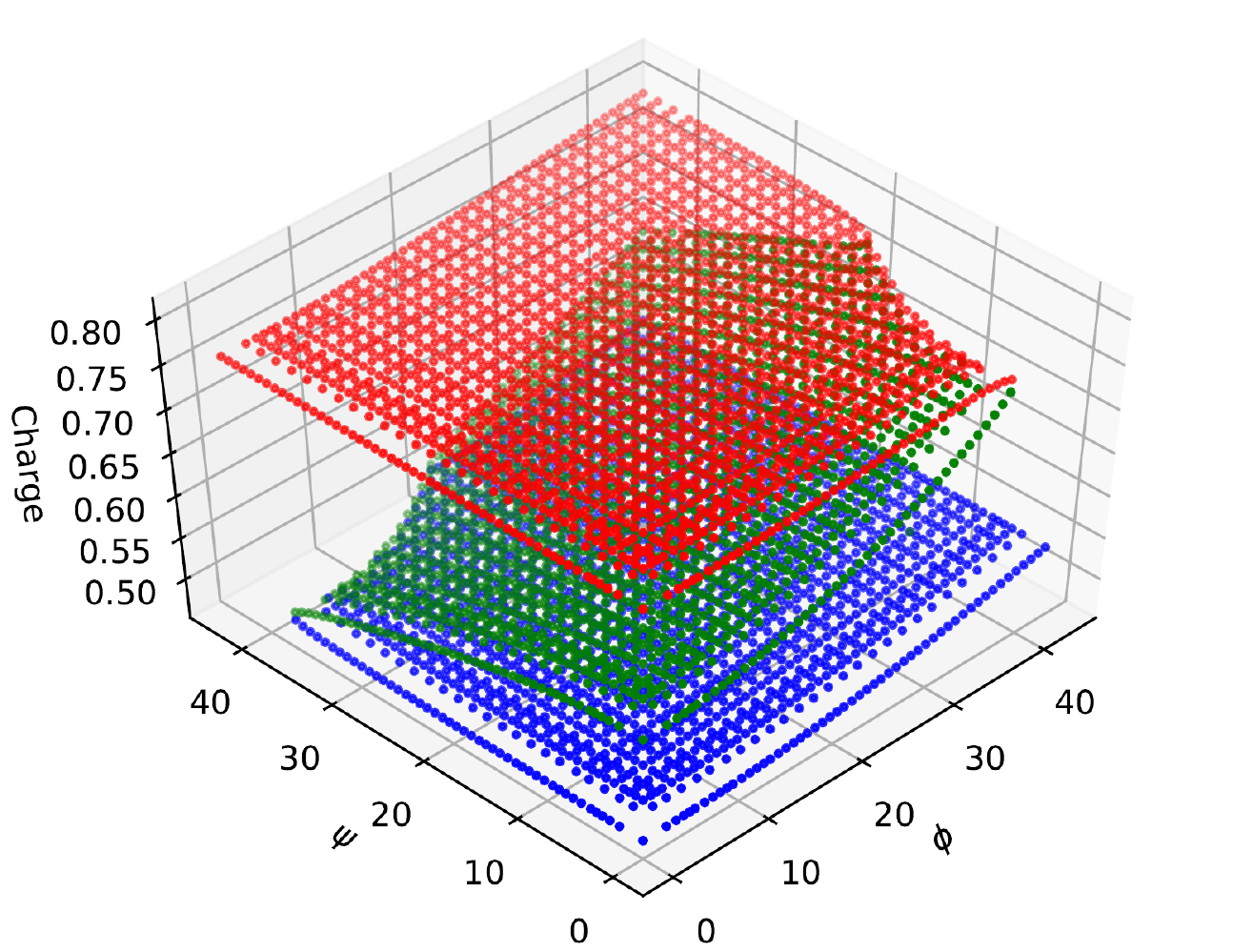}}\\
 \subfloat{\includegraphics[width=0.8 \linewidth]{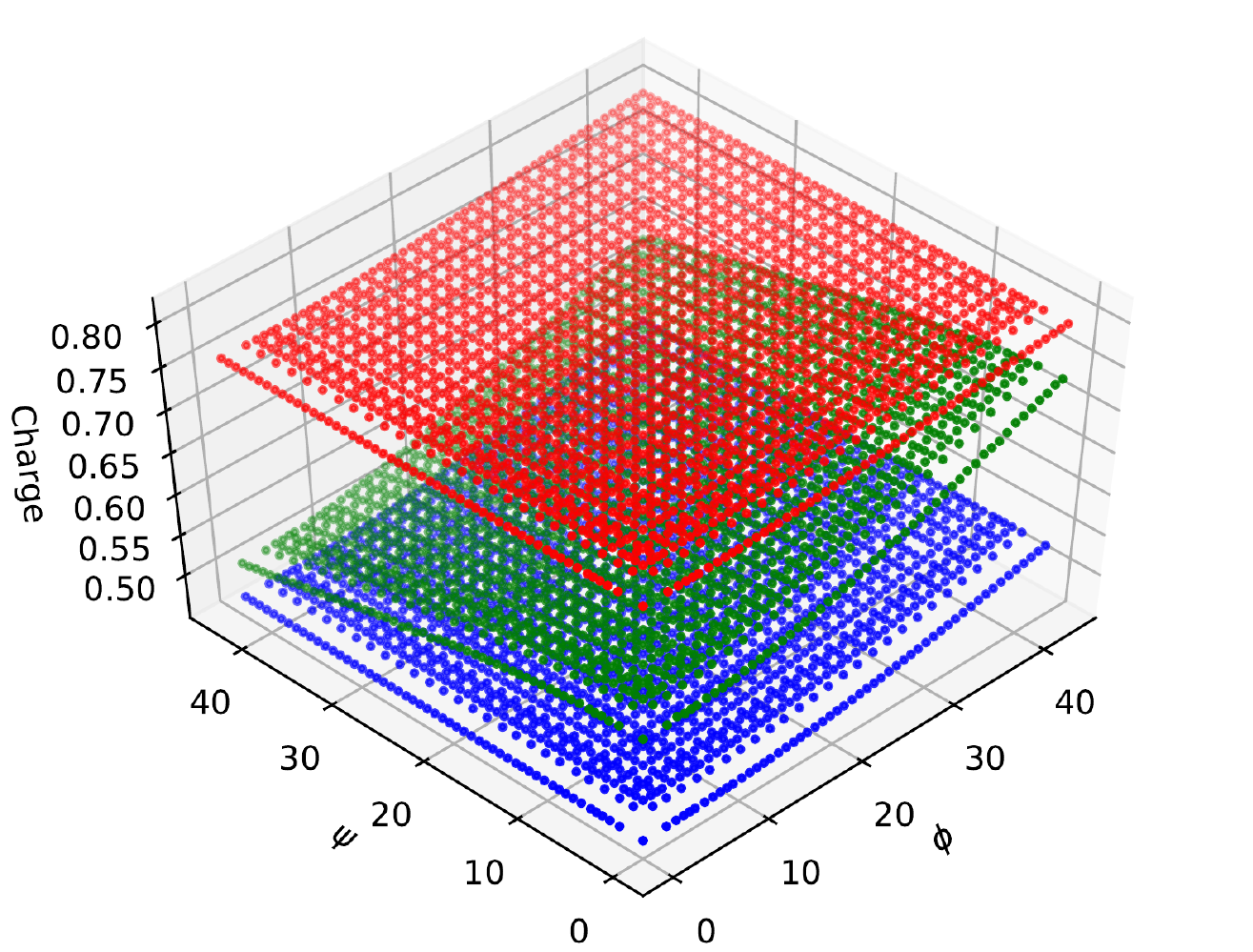}}
 \caption{\textit{top:} Mulliken charges on C1 for the D (red), A (blue) and E (green) SCF states of co-planar \ce{C7H6F4^.+} showing coalescence of states. \textit{bottom:} Mulliken charges on C1 for the D, A and E SCF states of staggered \ce{C7H6F4^.+} where no coalescence occurs.}
 \label{fig:C7H6F4_surfc}
\end{figure}

There exists a higher energy conformer of \ce{C7H6F4^.+} with the terminal \ce{CF2} groups staggered by $120\degree$ when viewed along the C1--C7 axis. With the terminal groups in this orientation, there are local geometric minima for both the D and A states, each of which has a different sets of $(\psi, \phi)$ values.
  In this conformation, the interaction between the D and A states is weaker, and coalescence of states does not occur in the investigated geometry space. The Mulliken charges on C1 for both this staggered conformer and the coplanar \ce{C7H6F4^.+} molecule for each of the A, D and E states are given in figure \ref{fig:C7H6F4_surfc}.\\

Figure \ref{fig:C7H6F4_surfc} emphasizes the quasidiabatic behavior of the D and A states in two-dimensions with roughly constant charges throughout the explored reaction space. This is in stark contrast to the behavior of the E state which changes its physical character significantly with molecular geometry. Thus while some SCF solutions appear to behave quasidiabatically, others are highly non-diabatic. Figure \ref{fig:C7H6F4_surfc} also suggests that the diabatic behavior of the D and A states becomes less pronounced in the vicinity of the line of coalesence.
This is in contrast to the behavior of the staggered conformer where no coalescence occurs and the donor and acceptor states thus maintain their quasidiabatic behavior throughout the investigated geometry space.

\subsection{Alizarin-Titanium SCF States}
\label{ss:alizarin_SCF}

\begin{figure}[h]
\hfill
 \subfloat{\adjincludegraphics[width=0.450 \linewidth,trim={0 0 0  {0.15 \height}}]{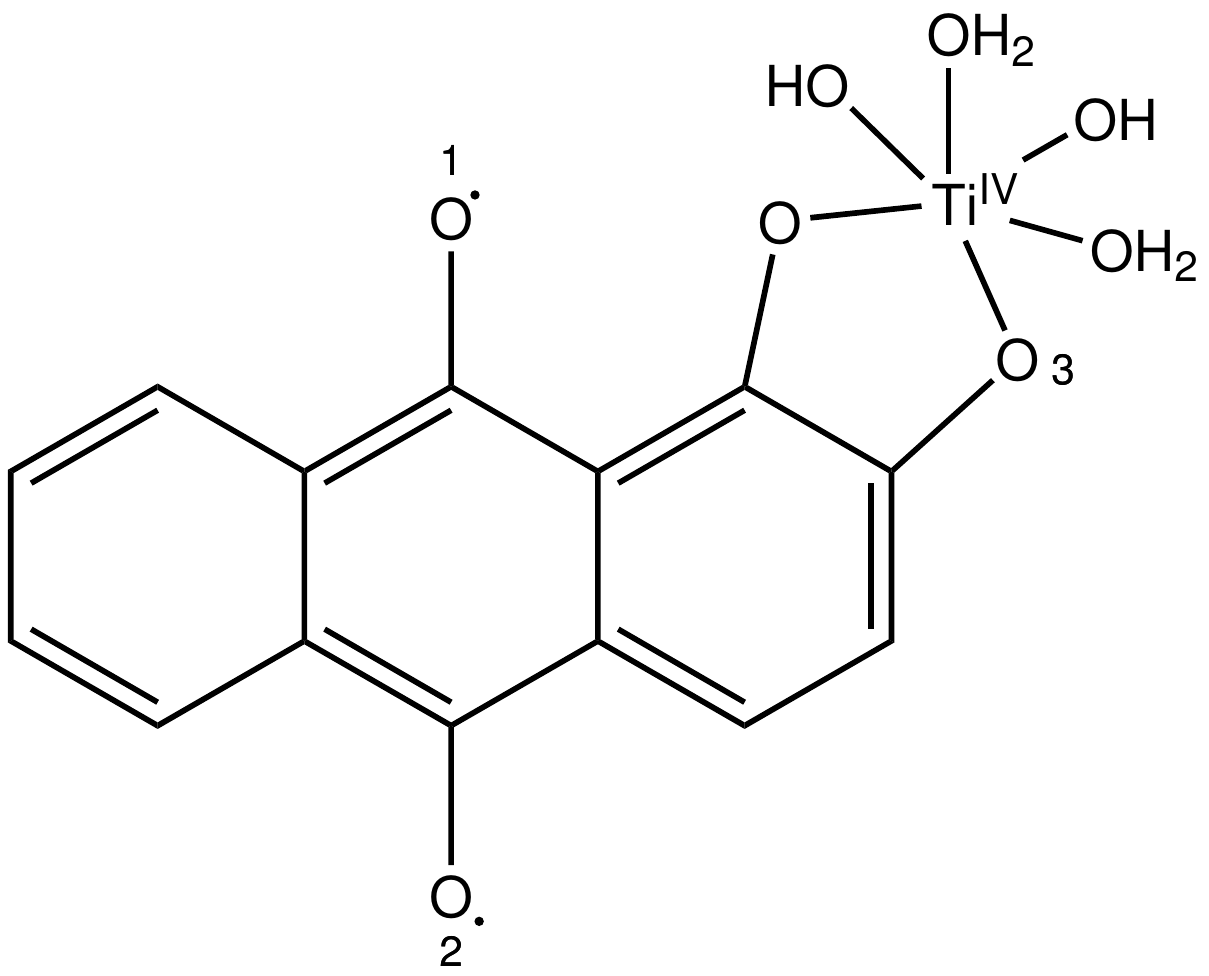}}
\hfill
 \subfloat{\adjincludegraphics[width=0.450 \linewidth,trim={0 0 0 {0.15 \height}}]{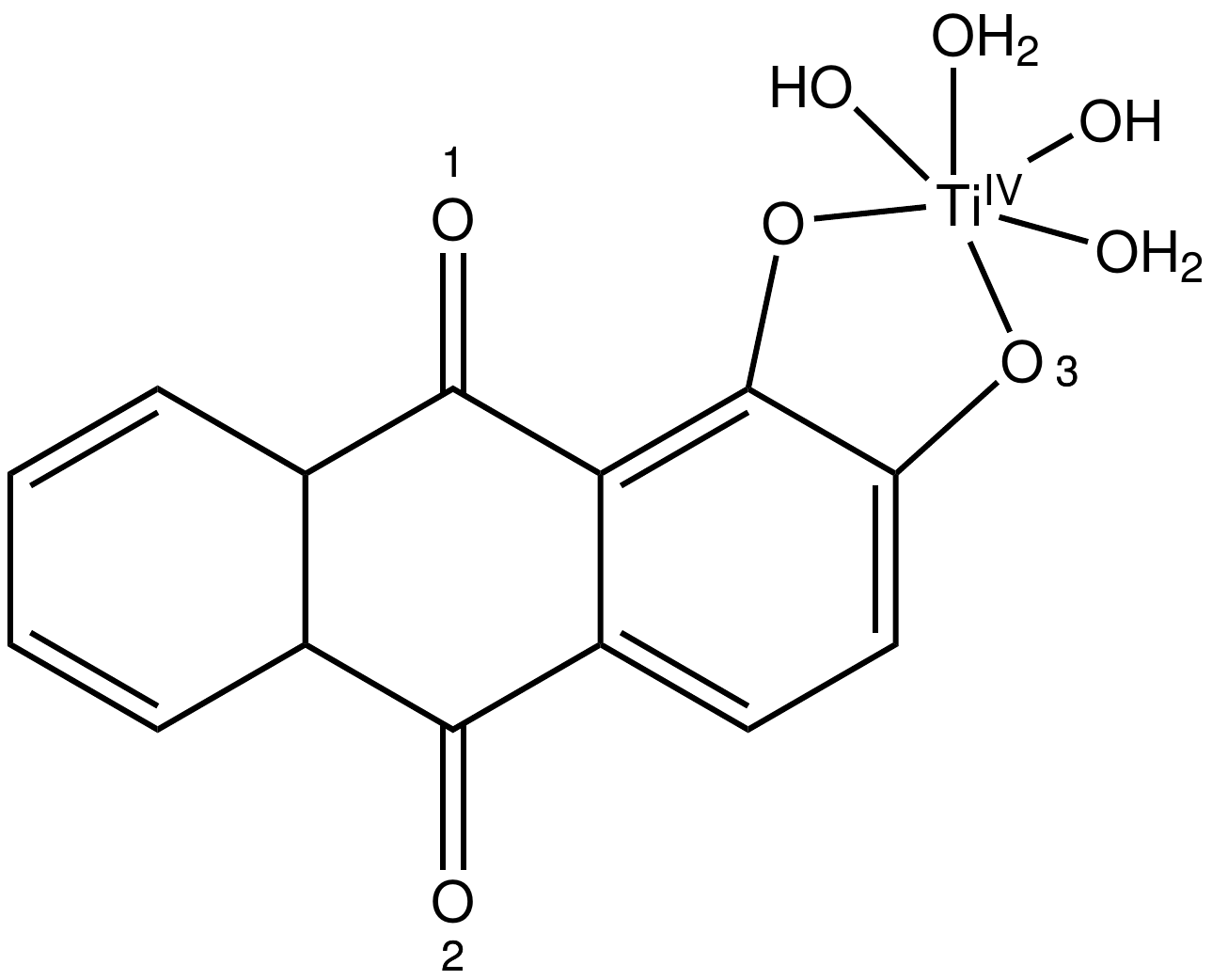}}
\hfill
\hfill 
\caption{Structure of the alizarin-titanium complex illustrating the donor di-radical ground state, D1 (\textit{left}), and quinone-like first excited state, D2 (\textit{right}). In the corresponding A states, an electron has been transferred from alizarin to Ti. This leaves a partial positive charge delocalized over alizarin with the largest contribution on O3.}
 \label{fig:structure_aliz}
\end{figure}

Encouraged by the results of the previous section and in order to show that the general principles investigated for \ce{C7H6F4^.+} also apply to more complex and asymmetric systems, we investigate the mechanism of electron transfer in a neutral \ce{Ti(OH)2(OH2)2}-alizarin complex (Figure \ref{fig:structure_aliz}) previously investigated by Duncan and Prezdho \cite{Duncan2005a}, working in a 6-31G* basis.\\

In recent years, Grätzel-type solar cells \cite{ORegan1991} and other dye-sensitized cells have gathered significant interest as a lower-cost alternative to conventional silicon-based solar cells \cite{Schwarz2000}.
Photoexcitation of a dye leads to injection of an electron into a semiconductor layer followed by reduction of the dye by an electrolyte-containing solution \cite{Hagfeldt2010}.
The basic principles of these cells have previously been modelled using computationally simpler systems such as the present titanium(IV)-alizarin system used as a model for a \ce{TiO2}-alizarin Grätzel type cell. In this system, the electron transfer occurs from the alizarin ligand to the Titanium atom, reducing it from \ce{Ti^{IV}} to \ce{Ti^{III}}. 
The donor electronic states of such isolated systems have been found both experimentally \cite{Wang2003} and computationally \cite{Duncan2005a} to be similar to those observed for the bulk materials, justifying their use as models of electron transfer.\\

An initial structure for the complex is generated using a na\"\i ve SCF geometry optimization (UHF with $M_S=0$) followed by a metadynamics calculation to identify states of interest. A Localized Orbital Bonding Analysis (LOBA)\cite{Thom2009b} is used to classify the identified solutions according to the oxidation state of the titanium atom. \ce{Ti^{IV}} states are labelled `D' and \ce{Ti^{III}} states `A' in accordance with the nomenclature used in the preceding section. Specific SCF geometry optimizations are carried out for each of the low energy D and A states in order to identify the donor and acceptor SCF ground states and their equilibrium geometries.\\

An NBO analysis \cite{Glendening2012} indicates that for both donor and acceptor geometries, the SCF ground state corresponds to a di-radical with spin-densities of 0.425 and -0.446 on O1 and O2 respectively for the donor ground state and 0.464 and -0.528 for the acceptor ground state. These are labelled D1 and A1. For both geometries, the first excited state, labelled D2 and A2, corresponds to a canonical quinone structure with approximately zero spin density on each of these two oxygens. These findings further support the notion of SCF metadynamics solutions as physically intuitive diabatic states even for larger systems. The second excited state for both donor and acceptor equilibrium geometries have intermediate spin densities of 0.203 and -0.165 for the donor state (D3) and 0.215 and -0.330 for the acceptor state (A3) on O1 and O2. Each D1-3 and A1-3 state also has a degenerate spin-flipped state which will be used in NOCI calculations.\\

\begin{figure}[h]
 \centering
\subfloat{\adjincludegraphics[width=0.8 \linewidth,trim={{0.05 \width} {0.30 \height} {0.05 \width}  {0.32 \height}}, clip=true]{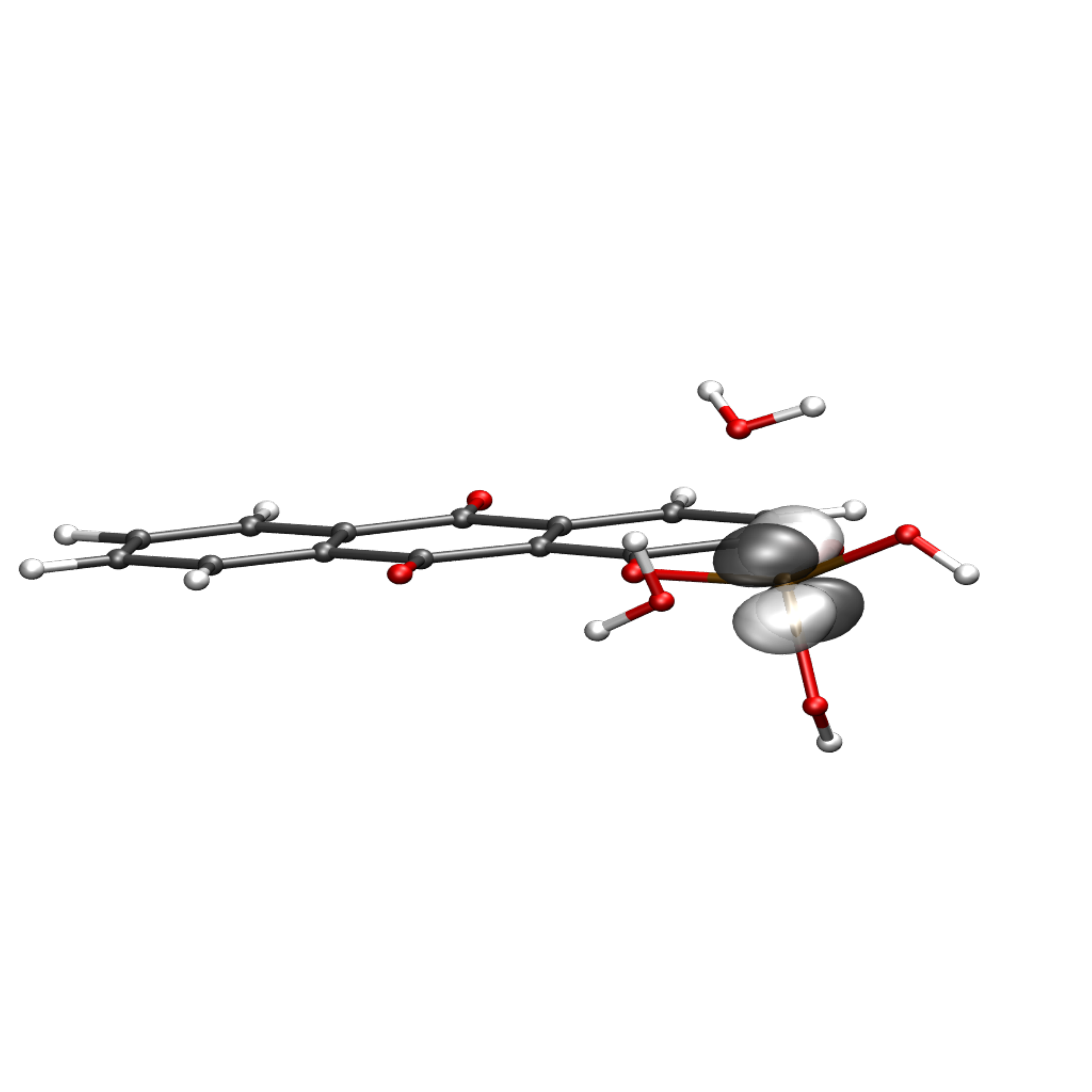}}
\\
\subfloat{\adjincludegraphics[width=0.8 \linewidth,trim={{0.05 \width} {0.30 \height} {0.05 \width}  {0.32 \height}}, clip=true]{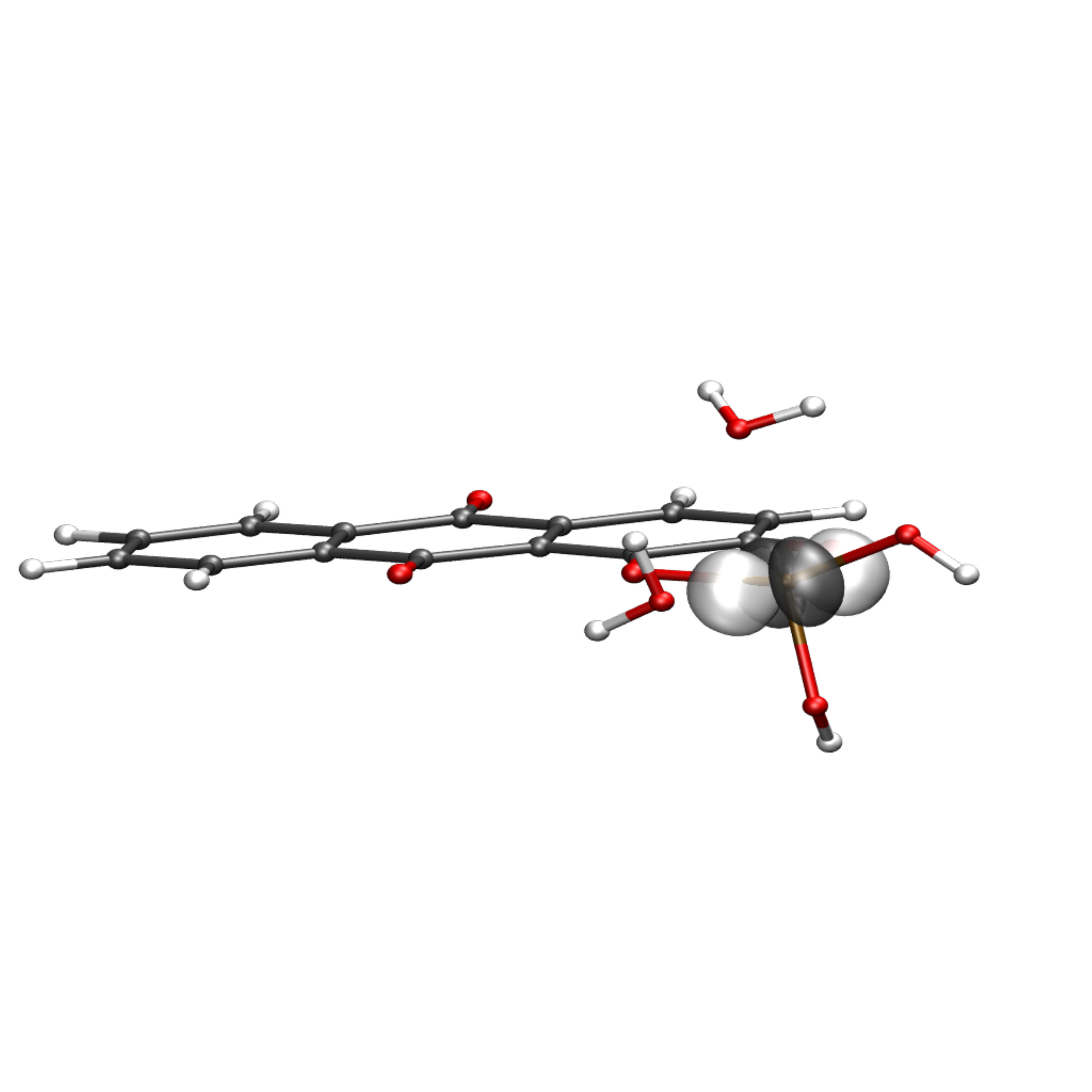}}
 \caption{The single localized d-orbital on Ti for the A1 state (\textit{top}) and a pseudo-orthogonal SCF state with different d occupancy  (\textit{bottom}) at the MECP geometry. Only the states with similar d-occupancy to A1 have been used in the remainder of the paper.}
 \label{fig:Ti_d_orb}
\end{figure}

In addition to the A1-3 states, there are several SCF solutions with alizarin electronic structures similar to A1-3, but which occupy a differently oriented d-orbital.
We have found that states with different d-occupancies do not interact significantly in NOCI, so we have restricted our further investigations to the set of states with similar d-occupancies to A1-3 (as determined from an NBO analysis\footnote{Although the D1-3 and the E1-3 states have no fully localized d-orbitals on Ti, the NBO analysis still indicates a similarly oriented d-orbital projection to the A1-3 states. In practice the D1-3 and E1-3 can be obtained by following the A1-3 states over different geometries to a point of coalescence.}). A pair of such pseudo-orthogonal d-states are shown in figure \ref{fig:Ti_d_orb}.  
 All energies are given relative to the NOCI ground state energy at the donor equilibrium geometry.\\

In order to investigate the behavior of the SCF solutions in more detail, we identify an approximate MECP between the D1 and A1 states using a quasi-Newton optimization. We generate an approximate reaction trajectory by interpolating between the MECP and both donor and acceptor geometries, tracking all of the D1-3 and A1-3 states along the trajectory. Here, it is worth noting that in the larger basis of 6 SCF states, the A1-D1 MECP is not likely to represent a transition state of the NOCI energy for electron transfer. However, we envision that it will be similar in geometry to a point on the minimum energy reaction trajectory and that forcing the interpolation through the MECP will approximate this trajectory more closely than a simple interpolation between donor and acceptor equilibrium geometries.
\\

\begin{figure}[h]
 \centering
 \subfloat{\includegraphics[width=1 \linewidth]{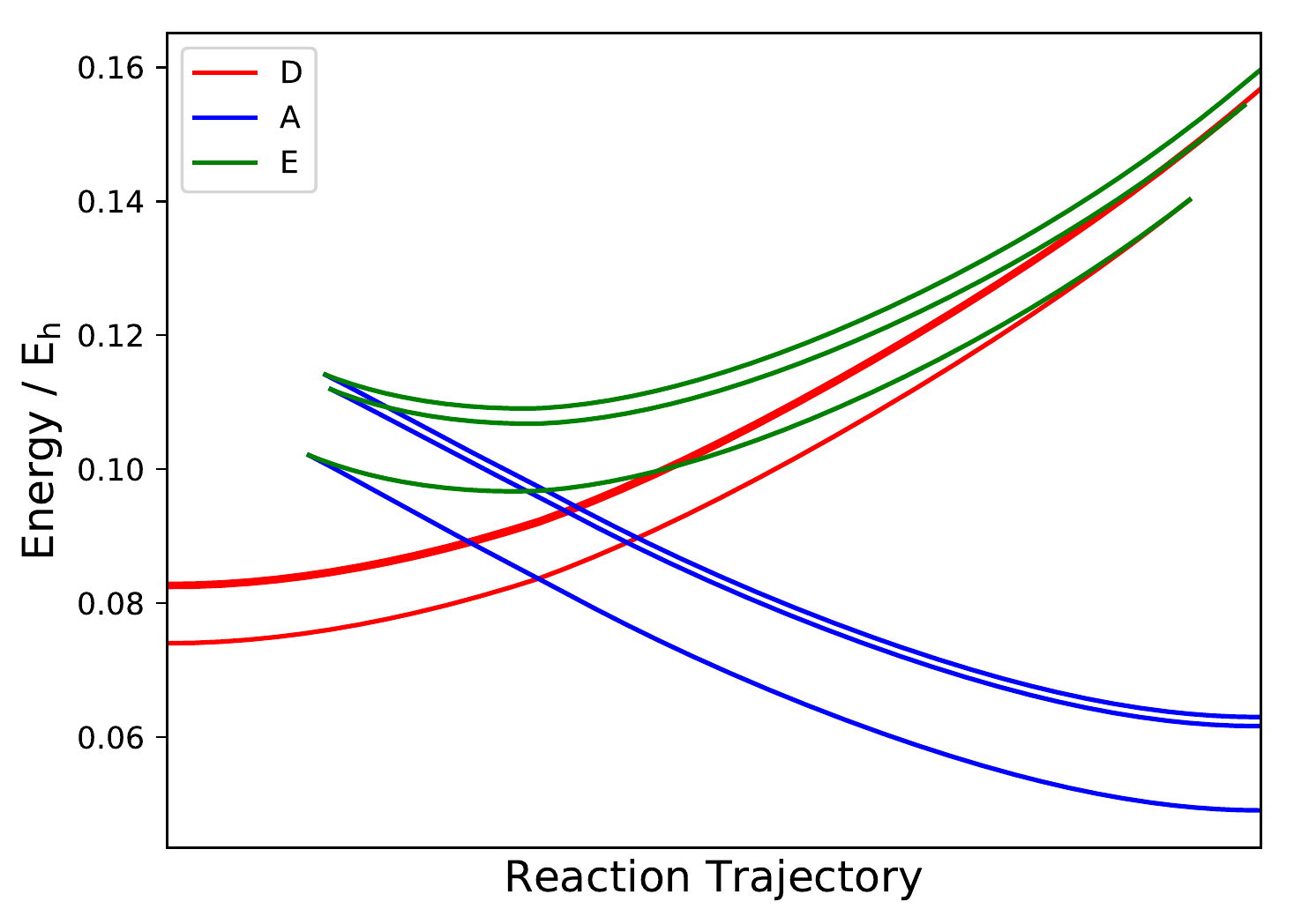}}
 \caption{Energies of the three lowest energy D, A and E states illustrating coalescence of states towards the ends of the reaction trajectory}
 \label{fig:alizarin_coalescence}
\end{figure}

As expected from previous results, disappearance of states is observed along the reaction trajectory with states A1-3 disappearing near the donor geometry and D1-2 disappearing near the acceptor geometry.
Surprisingly, we find that each of the D$n$-A$n$ pairs coalesces with the same excited SCF state, and we denote these excited states E1, E2 and E3. Thus E1 coalesces with D1 near the acceptor geometry and with A1 near the donor geometry while E2 and E3 coalesce with A2 and A3 respectively near the donor geometry, and E2 coalesces with D2 near the acceptor geometry  (Figure \ref{fig:alizarin_coalescence}).\\

We note that the D3 and E3 states do not coalesce within the reaction trajectory. This is similar to the staggered conformation of \ce{C7H6F4} (figure \ref{fig:C7H6F4_surfc}) where no coalescence is observed. However, coalescence of D3 and E3 may still occur at geometries not included in the proposed reaction trajectory, as would e.g. be the case for the A and E state of staggered \ce{C7H6F4} at the donor geometry if distorted towards the eclipsed donor geometry where these states are known not to exist.\\


The observed coalescence is intriguingly similar to  our previous observations for eclipsed \ce{C7H6F4^.+} where the non-diabatic E state coalesces with the quasidiabatic D and A states at opposite ends of the reaction trajectory.
We believe that this may be a general feature of the HF energy functional where minima represent quasidiabatic states which are connected by non-diabatic saddlepoints. In agreement with this hypothesis, a stability analysis of the SCF solutions reveals that the A and D states are stable whereas the Hessians for the E states have a single negative eigenvalue\cite{Cizek1967}.\\

Adjacent minima can be interpreted as electronic states which differ by a single electron and the intervening saddle point as a single-determinant approximation to a linear combination of the two minima. When propagating the system in certain directions in geometry space, a minimum may gradually flatten out, bringing it closer to the saddle point. Propagation of the system in the opposite direction can lead to the other minimum of the pair converging towards the saddle point.\\

\begin{figure}[h]
 \centering
	\subfloat{\adjincludegraphics[width=1 \linewidth, trim={{0.001 \width} {0.01 \height} {0.001 \width} 0}, clip=true]{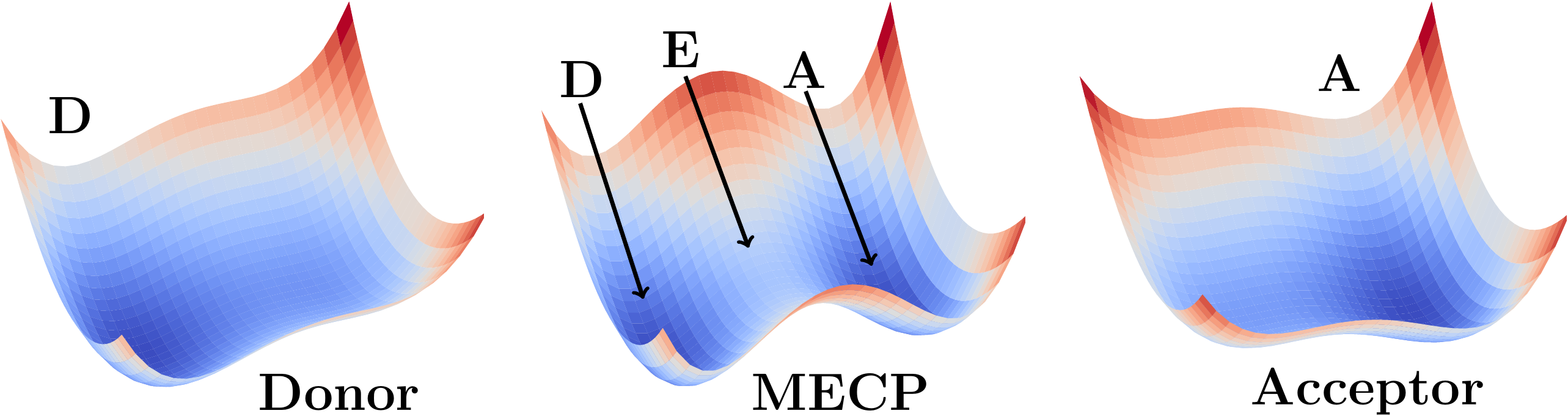}}
 \caption{Two-dimensional model of the behavior of the HF functional with changing molecular geometry. At the MECP (\textit{center}), the D and A electronic states correspond to minima of the energy functional while E is a saddle point. Near the donor equilibrium geometry (\textit{left}), the A and E states coalesce and only the D state is recovered as a solution to the HF equations. Near the acceptor equilibrium geometry (\textit{right}), the D and E states coalesce.}
 \label{fig:coeffs}
\end{figure}

In chemical terms, this corresponds to the relative contributions of the two diabats, D and A, to the single-determinant adiabat E changing with the geometry of the system. At certain points in geometry space, a minimum and a saddle-point coaslesce as they both cease to be stationary points and thus no longer appear as solutions to the Hartree--Fock equations.
This model of the behavior of the HF stationary points with changing geometry is illustrated in Figure \ref{fig:coeffs} using the nomenclature of the electron transfers described in the present paper. However, we expect that similar behavior will be observed for any process involving a change in molecular geometry.\\

This behavior of the present SCF solutions is similar to the SCF solutions of \ce{H2} where the RHF solution is a saddle point between the two UHF solutions at longer bond lengths but coalesce with the UHF solutions at shorter bond lengths. These similarities further support the hypothesis that the coalescing minima will reappear as holomorphic states after the point of coalescence.

\subsection{Alizarin-Titanium NOCI States}
\label{ss:alizarin_NOCI}

The SCF energies and energy differences are not expected to be good approximations to the real alizarin energy levels since these SCF solutions are spin contaminated and do not recover any of the electron correlation of the system.
We expect that by including a sufficient number of SCF states in a NOCI calculation, most of the static correlation of the system can be recovered and reasonable approximations to vertical excitation energies and relative energies of donor and acceptor states may be extracted.\\

\begin{figure}[h]
\centering
 \subfloat{\includegraphics[width=1 \linewidth]{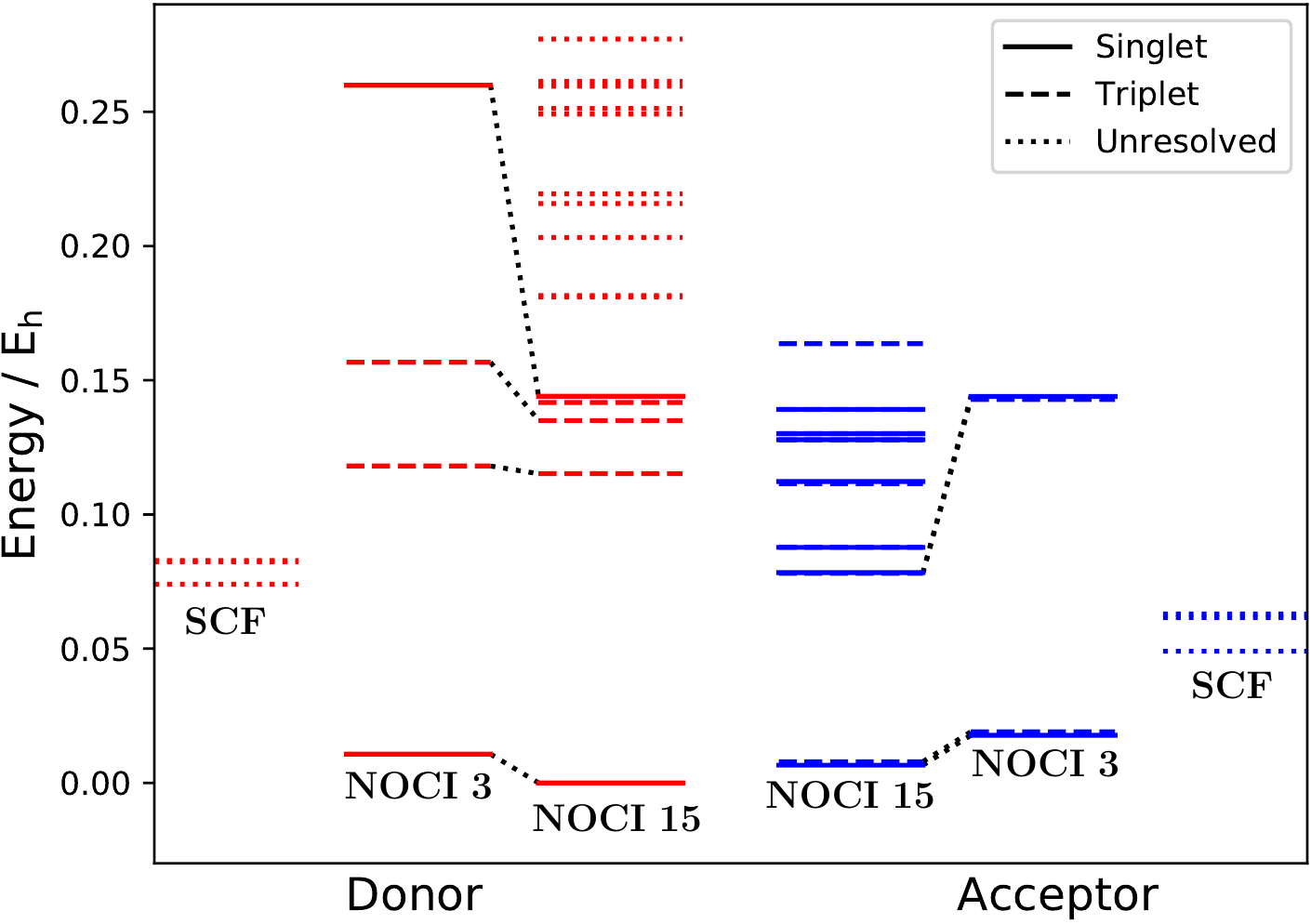}}
 \caption{Energies of the three lowest energy donor (D1-3) and acceptor (A1-3) SCF states; the four lowest energy states from a 3-SCF NOCI calculation; and the 15 lowest energy states from a 15-SCF NOCI calculation. NOCI at the donor equilibrium geometry was carried out using +IV electronic states (red), and NOCI at the acceptor equilibrium geometry was carried out using +III electronic states (blue). Black dotted lines indicate correspondence between the two lowest singlet and two lowest triplet states for a pair of NOCI calculations}
 \label{fig:NOCI_ens}
\end{figure}

We therefore perform a set of NOCI calculations on the 3 +IV SCF states with the lowest energies (D1-3) and their spin-flipped counterparts at the donor geometry, and the 3 +III states with the lowest energies (A1-3) and their spin-flipped counterparts at the acceptor geometry. A similar set of calculations is performed including more excited states, amounting to 15 spin-flipped pairs of states in each calculation. As illustrated in Figure \ref{fig:NOCI_ens}, this results in a significant lowering of the electronic energies. The resulting NOCI states are categorized according to spin multiplicity by inspection of the NOCI eigenvector coefficients for pairs of spin-flipped SCF states.\footnote{In Q-Chem, spin-flipping is achieved by swapping the spatial wavefunctions of a Slater determinant, and in this case the NOCI wavefunctions that consist of in-phase combinations of pairs of spin-flipped SCF states have symmetric spatial parts and must therefore have antisymmetric spin-parts, rendering them singlet states. Similarly, the NOCI wavefunctions consisting of out-of-phase combinations of pairs of spin-flipped SCF functions are triplet states. For some of the high-energy NOCI states, not all pairs of spin-flipped coefficients have equal magnitudes and these are therefore not classified according to spin multiplicity. This behavior is likely due to linear dependencies in the set of SCF states used for the calculation.}
\\

The energy difference between the two lowest energy donor singlet states is $0.1440367 \, \mathrm{E_h}$ corresponding to 31,600 cm\textsuperscript{-1}. This is comparable to the spectroscopically determined excitation energy for the alizarin-\ce{TiO2} complex of 20,100 cm\textsuperscript{-1  }\cite{Huber2000, Rajh2002}. It may be expected that the discrepancy is due to a combination of the inability of the 6-31G* basis to describe large d-complexes accurately, and the fact that using a basis of only 15 SCF solutions is likely to give a significantly better description of the ground state than excited states, leading to an overestimation of excitation energies.\\

\begin{figure}[h]
\centering
 \subfloat{\includegraphics[width=1 \linewidth]{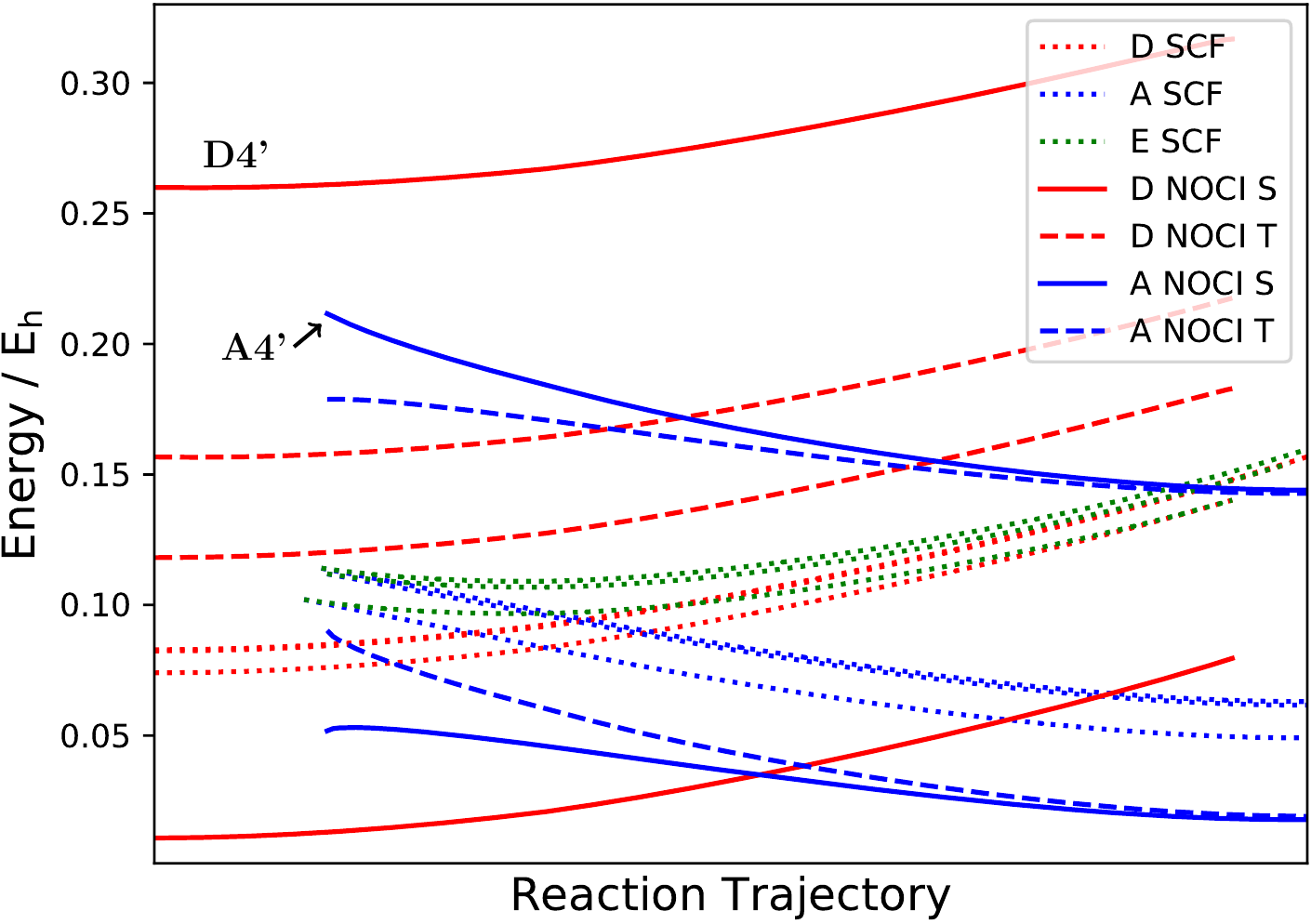}}
 \caption{Energies of the four lowest donor and acceptor NOCI states together with the three lowest energy D, A and E SCF states. Solid line: singlet (S), dashed line: triplet (T).}
 \label{fig:alizarin_NOCI}
\end{figure}

Figure \ref{fig:NOCI_ens} also shows that including only three states in the NOCI calculation appears to be a reasonable approximation to the energies and spin multiplicities of the higher level NOCI calculations, especially for the low energy states.
We expect that by performing NOCI on a set of +IV or +III SCF states that behave quasidiabitally, and the energies of which follow similar trends when varying the molecular geometry, we generate sets of quasidiabatic +IV and +III NOCI states.
NOCI calculations were therefore performed on the three pairs of spin-flipped donor states with lowest energy (D1-3) throughout the part of the reaction trajectory where they all exist, and similarly for the acceptor states, in order to investigate the behavior of the diabatic states of the system (Figure \ref{fig:alizarin_NOCI}). These quasidiabatic NOCI states are denoted D$n$' and A$n$' in accordance with the nomenclature used for diabatic SCF states. \\

Figure \ref{fig:alizarin_NOCI} can be interpreted as a non-adiabatic picture of the electron transfer and is qualitatively consistent with previous models\cite{Huber2000} suggesting that an electron is excited from the alizarin ground state to the first excited singlet state (in this case D4') followed by transfer to an acceptor singlet state which is likely to be the first excited singlet acceptor state (A4'). Since this acceptor state has an additional electron on Ti compared to the donor state, this corresponds to the electron being transferred to titanium and is followed by a barrierless relaxation to the acceptor equilibrium geometry. The detailed behaviour of the A4' state is not expected to be similar to that for an actual Grätzel-type cell since only the donor states of model systems such as ours behave similarly to the bulk system.
This requires a crossing of the D4' and A4' energy curves near the donor equilibrium geometry, which would happen after the point of coalescence of the diabatic A and E states. We hypothesize that if the corresponding holomorphic states could be located and tracked, diabatic NOCI states formed from these would show such a crossing, thus completing the picture of the electron transfer.\\

\begin{figure}[h]
\centering
 \subfloat{\includegraphics[width=1 \linewidth]{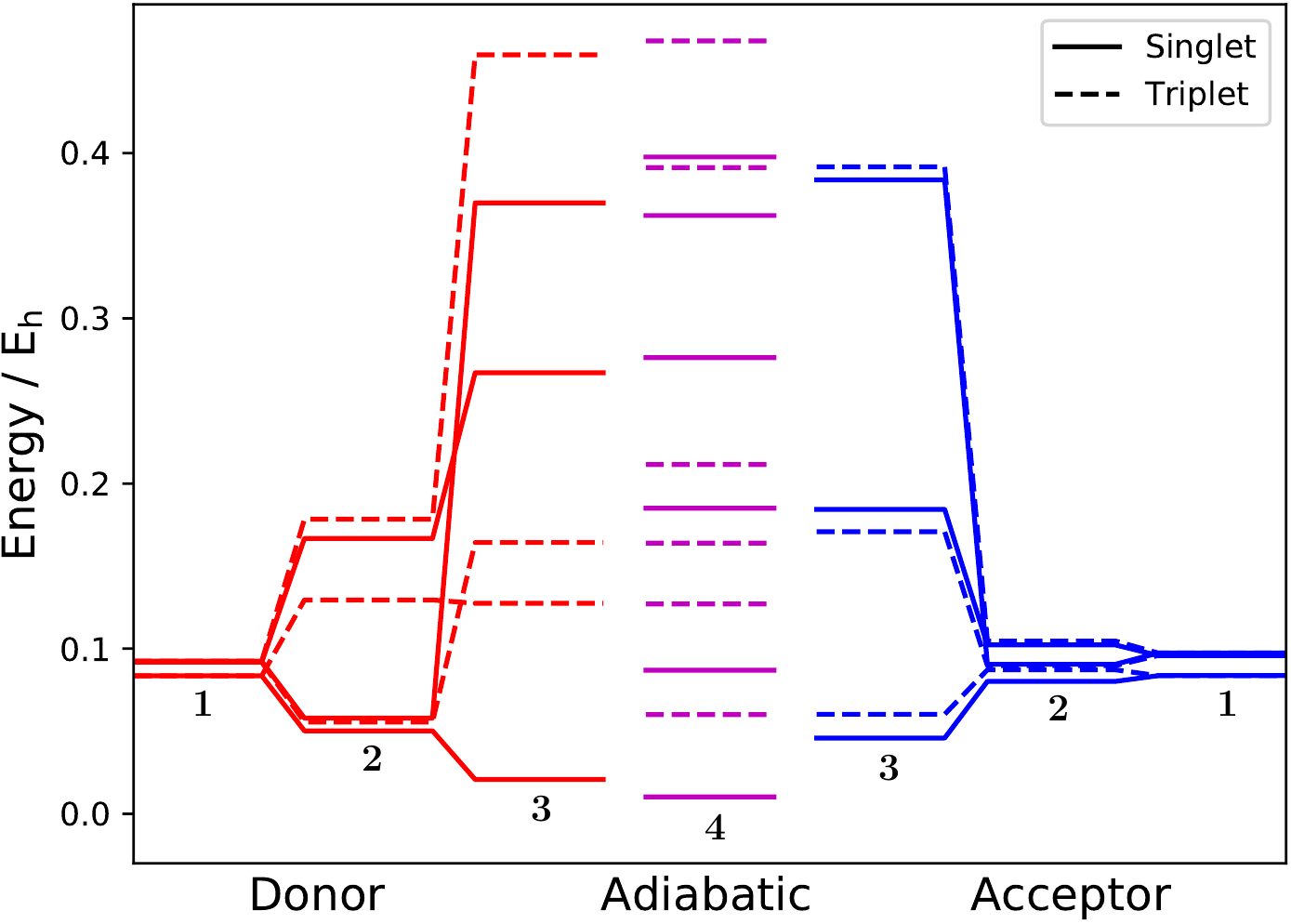}}
 \caption{Energies of (\textit{1}) SCF states, (\textit{2}) NOCI between pairs of spin-flipped SCF states, (\textit{3}) NOCI between sets of three pairs of spin-flipped SCF states, and (\textit{4}) NOCI between all six pairs of spin-flipped SCF states.  Solid lines: singlet states. Dashed lines: triplet states. Calculations were performed at the MECP geometry.}
 \label{fig:MECP_seq}
\end{figure}

The quasidiabatic NOCI states in Figure \ref{fig:alizarin_NOCI} can be further interacted with each other to give a set of adiabatic NOCI states.
Figure \ref{fig:MECP_seq} illustrates such a set of hierachical NOCI calculations at the MECP geometry for both the donor and acceptor states and shows how the energies change when more states are included in a NOCI calculation. The first sets of 3 states illustrate the pure SCF states which are then interacted separately with their spin-flipped counterparts to generate two sets of 6 spin-resolved states, one for the donor states and one for the acceptor states.
The third sets of donor and acceptor states in Figure \ref{fig:MECP_seq} are the quasidiabatic NOCI states resulting from interacting the spin-resolved donor or acceptor states. Finally, the quasidiabatic donor and acceptor NOCI states are interacted to generate 12 adiabatic NOCI states.\\

\begin{figure}[h]
\centering
 \subfloat{\includegraphics[width=1 \linewidth]{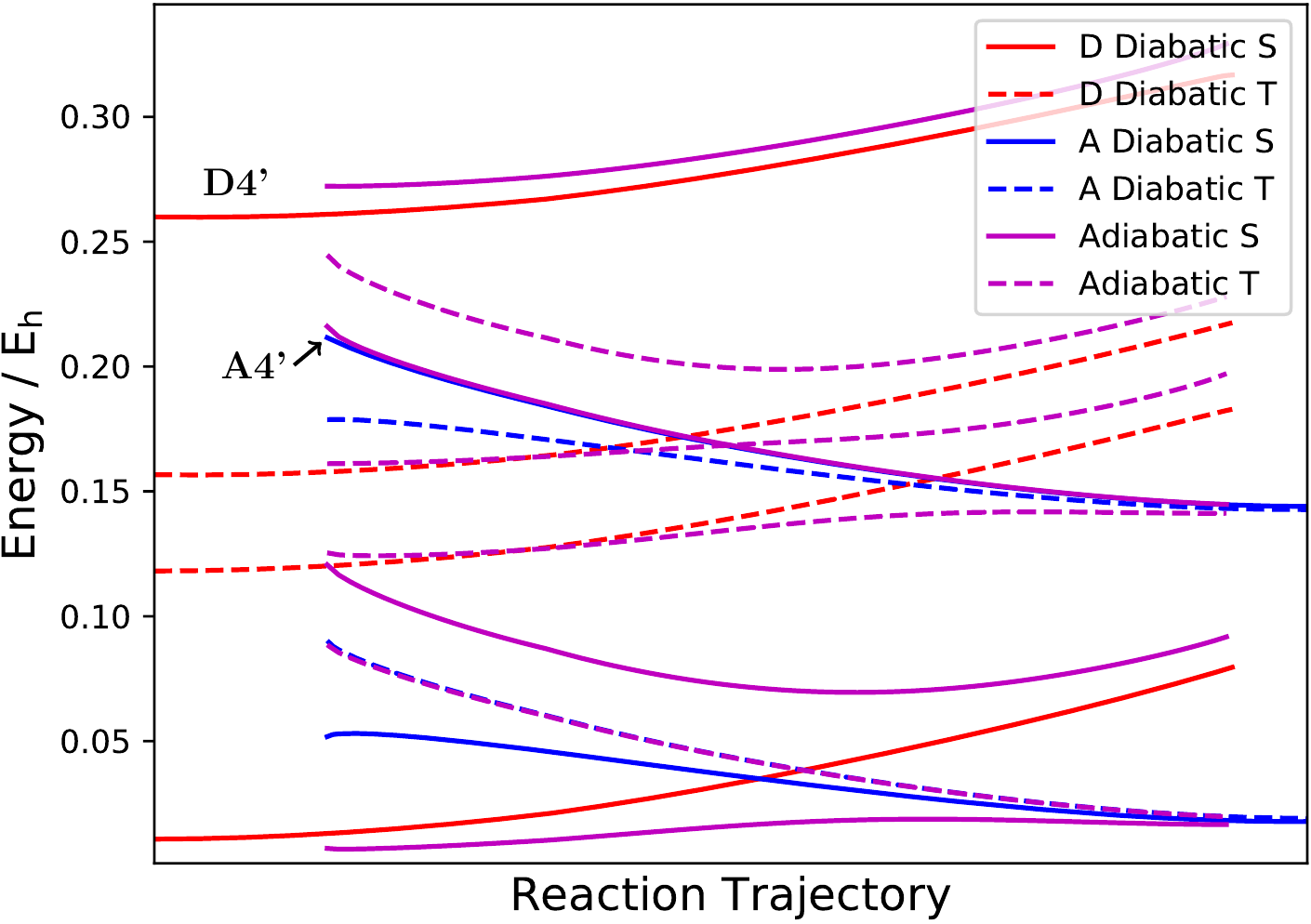}}
 \caption{Energies of the four lowest energy diabatic donor and acceptor NOCI states together with the eight lowest adiabatic NOCI states. Solid line: singlet (S), dashed line: triplet (T).}
 \label{fig:alizarin_NOCI_tot}
\end{figure}

As expected, the donor singlet states only interact with acceptor singlet states and similarly for the triplet states. This has significant consequences for the behavior of the adiabatic NOCI states with e.g. avoided crossings only being observed between states of the same spin multiplicity when tracking the adiabatic NOCI states across the reaction trajectory  (Figure \ref{fig:alizarin_NOCI_tot}).\\

The adiabatic ground state is a singlet state and describes a hypothetical thermal adiabatic electron transfer from alizarin to titanium. The activation energy for this adiabatic transfer is 21.03 kJ/mol which is significantly higher than \textit{kT} at room temperature (2.475 kJ/mol) and thus prohibitive of thermally induced electron transfer. This may explain why the alizarin-to-titanium electron transfer must be light-induced and thus why alizarin is a useful molecule for constructing dye-sensitized cells.\\

In the current framework, a model of the electron transfer consistent with existing models instead involves local excitation of an electron from the D1' state to the D4' state on alizarin followed by transfer of the electron to the A4' state on titanium.  This is finally followed by a barrierless relaxation to the acceptor geometry to complete the electron transfer process.\\

This alizarin-titanium system also illustrates how the methodology described can be applied to more complex electron transfer processes. In this case, an extensive metadynamics calculation is performed for a trial geometry and each Hartree-Fock solution is categorized as belonging to the set of `donor' or `acceptor' states based on e.g. a Mulliken of LOBA analysis.\\

Once donor and acceptor states have been identified, they can be optimized to locate donor and acceptor equilibrium geometries. For these geometries, it is important to perform another metadynamics calculation to identify any states that may not exist for the trial geometry due to coalescence. The sets of `donor' and `acceptor' SCF states can then be used as bases for diabatic NOCI states which can be combined to generate an adiabatic picture of the electron transfer. The method is scalable for well-defined electron transfers where it is possible to categorize the low-energy SCF solutions.

\section{Conclusions}
In conclusion, the present investigations shed further light on the behavior of multiple SCF solutions and their ability to describe physically relevant systems. The results indicate that the SCF solutions corresponding to minima of the energy functional behave quasidiabatically and in many cases correspond to physically intuitive electronic states of the system. These can be used as a basis to describe non-adiabatic electron transfer, or they can be combined in a Non-Orthogonal Configuration Interaction calculation to yield adiabatic energy curves.\\

At certain points in geometry space, SCF minima have been found to coalesce with saddlepoints of the HF energy functional. The saddlepoints behave non-diabatically and can be interpreted as single-determinant approximations to linear combinations of pairs of SCF wavefunctions. The saddle points tend to coalesce with the higher energy state of the pair of minima at large energy separations. \\

By tracking excited SCF solutions and generating appropriate NOCI states for an alizarin-titanium system of relevance to the study of dye-sensitized solar cells, we were able to map out the electron transfer from alizarin to titanium and describe the electronic states involved in this process in more detail than has been achieved in previous studies using Density Functional Theory and Non-Adiabatic Molecular Dynamics\cite{Duncan2005a,Duncan2005b}.\\

Among other observations, we were able to classify the relevant electronic states according to spin multiplicity and thus shed further light on the alizarin photoexcitation process. This may contribute to an increased understanding of how this particular solar cell works and is likely to be applicable to other similarly-sized systems.\\

The alizarin-titanium system illustrates how the methodology can be extended to treat systems where the donor and acceptor states are poorly described by a single determinant. This is achieved by combining sets of Hartree--Fock donor or acceptor states to generate diabatic donor and acceptor NOCI states for a non-adiabatic description of the electron transfer process. These can be further interacted to give an adiabatic picture of the electron transfer.\\

In summary, we believe that using multiple SCF solutions and Non-Orthogonal Configuration Interaction to model electron transfer processes can be a useful tool for gaining a qualitative understanding of the electron transfer and the electronic states involved, as well as providing quantitative approximations to excitation energies and adiabatic interactions.\\

\section{Computational Methods}
All calculations were performed in Q-Chem 4.4.1\cite{Shao2015}.\\

SCF calculations were considered to have converged when the wavefunction error was less than $10^{-7}$. Lowering the threshold from this value was found to lead to lower quality SCF states with the resulting NOCI states having less smooth energy curves. All calculations were performed as Unrestricted Hartree Fock calculations.\\

Multiple Hartree-Fock solutions were found using SCF Metadynamics as described by Thom and Head-Gordon\cite{Thom2008} and implemented in Q-Chem. This involves performing a Hartree-Fock calculation using a modified energy functional that includes a Gaussian penalty function to avoid reconvergence to minima that have already been located.\\

Non-Orthogonal Configuration Interaction (NOCI) calculations were performed as described by Thom and Head-Gordon\cite{Thom2009} and implemented in Q-Chem. \\

Minimum Energy Crossing Points (MECPs) were identified by performing an optimization of the energy of the D state constrained to be degenerate with the A state. This was done my minimizing the lagrangian\\$L(\mathbf{R};\lambda)=E_D(\mathbf{R}) - \lambda (E_D(\mathbf{R})-E_A(\mathbf{R}))$\\ as described by Koga and Morokuma\cite{Koga1985}. In the case of \ce{C7H6F4+} this was done using the analytical Hessian. In the case of the larger alizarin complex, an approximate Hessian was used in a quasi-Newton optimization where the inverse Hessian is updated according to\\
$(H^{(k+1)})^{-1} = (H^{(k)})^{-1} + \dfrac{\mathbf{u} \mathbf{u}^{\top}}{\mathbf{u}^{\top} \mathbf{\gamma}^{(k)}}$\\
$\mathbf{u} = \begin{pmatrix} \delta_\mathbf{R} \\ \delta_{\lambda}  \end{pmatrix} - (H^{-1})^{(k)} \gamma^{(k)}$\\
$\gamma^{(k)} = \nabla_{\mathbf{R}, \lambda}^{(k+1)} - \nabla_{\mathbf{R}, \lambda}^{(k)}$\\
The calculations were considered to have converged when the square norm of the total gradient vector with respect to all nuclear coordinates and the Lagrangian was less than 0.005.

\begin{acknowledgement}

AJWT thanks the Royal Society for a University Research Fellowship (UF110161) and Magdalene College for project funding for RLB and KTJ.

\end{acknowledgement}

\begin{suppinfo}

Geometries of all stationary points investigated.\\
SCF and NOCI electronic energies of all stationary points investigated.

\end{suppinfo}


\bibliography{mentest}

\end{document}